\documentclass[aps,prb,reprint,groupedaddress]{revtex4-2}
\usepackage{amssymb}
\usepackage{bm}
\usepackage{graphicx}
\usepackage{amsmath}
\usepackage{mathtools}

\graphicspath{{Figures/}}

\begin{document}

\title{Oscillations in the radiative damping of plasma resonances in a gated disk of 2D electron gas}

\author{D.A. Rodionov}
\affiliation{Kotelnikov Institute of Radioengineering and Electronics of Russian Academy of Sciences, Moscow, 125009 Russia}
\affiliation{Moscow Institute of Physics and Technology, Dolgoprudny, Moscow Region, 141701 Russia}

\author{I.V. Zagorodnev}
\email[]{zagorodnev@phystech.edu}
\affiliation{Kotelnikov Institute of Radioengineering and Electronics of Russian Academy of Sciences, Moscow, 125009 Russia}

\date{\today}

\begin{abstract}

We calculate the absorption spectra of a conductive disk in the presence of a metal gate. In our analysis, we use the conductivity described by the Drude model. We examine the frequency and line width of the absorption resonances associated with the excited plasma waves. They are estimated both numerically and analytically for the fundamental and axisymmetric plasma modes. The line width is determined by the collisional and radiative losses. In high mobility samples, the latter may dominate even when the resonant frequencies can be described by a standard quasistatic approximation, i.e., neglecting electromagnetic retardation. Placing a metal near the disk drastically affects the radiative broadening of the line width due to the interference of electromagnetic fields created by the disk and the metal charges. Bringing the disk in close proximity to the gate makes the total field almost vanish. In that case, the line width is defined solely by the collisional broadening. As the disk is moved away from the gate, the line width first shows a reduction since the plasmon transforms into plasmon-polariton, which is dressed with undamped electromagnetic field surrounding the disk. Then it increases and exhibits decaying oscillations, reaching the asymptotic value of the ungated disk.

\end{abstract}

\maketitle

\section{Introduction}

The Coulomb interaction between freely moving charged particles in solids results in collective oscillations --- plasma waves, or plasmons. In bulk metals, the plasmon frequency is almost independent of its wavevector and has typical values around 1 eV/$\hbar$ due to high electron concentration \cite{Pines1952}. By contrast, in a two-dimensional (2D) electron gas, it is usually much lower, exhibiting a square root dependence on the wavevector in the simplest quasistatic (or quasi-electrostatic) approximation \cite{Stern1967}. The gapless dispersion makes the 2D plasmons promising in optoelectronic applications, in particular, in the terahertz range \cite{Dyakonov1993,Knap2009}. Recently, they have been intently discussed with regard to graphene structures \cite{Grigorenko2012,Yao2018,Bandurin2019,Mitin2020}.

An essential feature of the 2D plasmons is that their frequencies can be controlled and smoothly tuned by changing the carrier concentration using a gate - a planar metal electrode in the vicinity of a 2D electron system. The gate screens the Coulomb interaction, which results in changing the long wavelength frequency dispersion to a linear one with group velocity less than the speed of light \cite{Chaplik1972}. According to the energy and momentum conservation laws, the plasmons cannot be excited by an electromagnetic plane wave. Hence, to excite plasmons 'under the light branch', metallic gratings \cite{Meziani2008}, near-field optical microscopy \cite{Nikitin2016}, or size-limited sample geometries like disks or strips \cite{Zarezin2021} are used. These confined structures are especially compelling in practical applications because they effectively compress and absorb light \cite{Alcaraz2018,Epstein2020,Mylnikov2022}. The disk geometry, in particular, is one of the simplest configurations for manufacturing as well as theoretical analysis. 

Plasmons in a conductive disk in the presence of a metal gate (gated disk) have been studied since 1985 \cite{Glattli1985,Fetter1986}. However, the primary focus has been on the plasma resonant frequencies in the 'standard' quasistatic regime. In that case, all effects are independent of the speed of light; therefore, the plasmon damping is determined solely by the electron collisional decay. In an infinite 2D electron system, it occurs when the plasmon frequency is much lower than the frequency of light at the same wavevector \cite{Zabolotnykh2021}. Sometimes it is believed that the same reasoning can be applied to a confined 2D electron system, in particular, for a disk, wherein the wave vector is inversely proportional to the radius \cite{Gusikhin2018}. It is expected that the quasistatic approximation is valid when the sample size is much smaller than the wavelength of the electromagnetic radiation.  However, in such systems, the plasmon damping is affected not only by collisions but also by electromagnetic radiation \cite{Mikhailov2004}. The excited plasmon radiates the electromagnetic wave and is damped by the radiation power, but the power depends on the interference of the radiated wave from the plasmon in the disk and the wave reflected from the gate, so does the radiative damping rate. It is essentially a non-quasistatic phenomenon that can be properly described only by taking into account the electromagnetic retardation effects. In high mobility samples, with extremely long electron scattering time \cite{Chung2021}, the radiation can dominate the collisions, even when the plasmon frequency is much lower than the frequency of light, i.e., seemingly, in the standard quasistatic regime. This particular case is the main focus of the present article. We will establish the limits of applying the quasistatic approximation in evaluating the line width.

The presence of a gate near a 2D electron system affects the radiation losses due to the interference of the electromagnetic fields from the disk and the metal. Indeed, the latter can be regarded as the fields created by the image-disk charges --- the mirror reflection of the actual disk of the metal plane, with the opposite currents and charges \cite{Batygin1964}. Thus it results in constructive or destructive interference depending on the distance between the disk and the gate, similar to the radiation of a dipole in the vicinity of a mirror \cite{Li2010}. Therefore, the plasmon radiative damping is strongly influenced by the gate-disk separation distance, resembling the Purcell effect for an oscillator (atom) in a cavity \cite{Haroche1989} but in an open classical system.

Inspired by recent experimental achievements \cite{Chung2021,Rosales2022,Muravev2017,Zoric2011,Gusikhin2018,Andreev2014,Andreev2015,Andreev2021}, this work is devoted to the analysis of the frequency and decay rate of the axisymmetric (with angular momentum $l=0$) and fundamental ($l=1$) plasma modes in a gated disk. In Sec.~\ref{sec:2}, we discuss the principal equations and their proper parametrization. In Sec.~\ref{sec:3}, we consider an exact solution in a fully screened limit when the gate and the disk are very closely spaced. In that case, the plasmons are nonradiative, even at frequencies beyond the quasistatic regime. At the same time, the plasmon damping is reduced with increasing retardation. Therefore, at some point, the radiative losses exceed the effect of collisions. In the following sections, we determine the condition restricting the application of the fully screened approximation in estimating the plasmon damping in high-mobility samples. In Secs.~\ref{sec:4} and~\ref{sec:5}, we calculate the absorption power spectra for the axisymmetric and fundamental modes. Following the Galerkin method, we seek a solution to Maxwell's equations, expanding the unknown components of the current density in a Taylor-like series. Cutting the series, we find the approximation for the current density and then calculate the absorption spectra. From these data, we determine the dependence of the frequency and line width of the absorption peaks on the distance between the disk and the gate. The absorption maxima indicate the excited plasma resonances, while their line widths relate to the damping of the plasma modes. The interference of the electromagnetic fields originated from the disk and the nearby metal can suppress or enhance the radiative decay and, therefore, the Q-factor of the system. For example, when the gate is extremely close to the disk, it suppresses the total radiation, narrowing the plasma resonance line and reducing the system response. On the other hand, as the disk is moved away from the metal, the line width oscillates as a function of the separation distance. The oscillation amplitude decays with increasing separation as the line width asymptotically approaches the value corresponding to the ungated system.

\section{\label{sec:2}Key equations}

In this section, we formulate the equations describing the response of a gated disk and discuss the solution method. Consider a conductive disk of radius $R$ in the plane $z=0$ at a distance $d$ above the gate, which is assumed to be an ideal infinite metal plane, and let ${\bf r}$ be the radial vector in the disk plane. We seek the response of the disk charges to the external electric field ${\bf E}^{ext}\left({\bf r}\right)e^{-i\omega t}$ with the oscillation frequency $\omega$. In the following analysis, all vectors are regarded as 2D vectors in the disk plane. The net incident electric field includes the original field of the source as well as that reflected from the gate. The total electric field ${\bf E}^{tot}\left({\bf r}\right)e^{-i\omega t}$ represents the superposition of the net external field and the field induced by the disk charges and their image counterparts, ${\bf E}^{ind}\left({\bf r}\right)e^{-i\omega t}$. Then, the current density in the disk can be expressed as
\begin{equation} \label{eq:Current}
	{\bf j}\left({\bf r}\right) = \sigma\left(\omega\right) {\bf E}^{tot}\left({\bf r}\right) = \sigma\left(\omega\right) \left[ {\bf E}^{ext}\left({\bf r}\right) + {\bf E}^{ind}\left({\bf r}\right) \right].
\end{equation}
We use the Drude conductivity $\sigma\left(\omega\right) = ne^2\tau/m\left(1-i\omega\tau \right)$, where $n$, $m$, and $\tau$ are the 2D concentration, the effective mass, and the carriers relaxation time, respectively. In fact, it is governed only by two intrinsic parameters --- the collisional damping rate, $\gamma = 1/\tau$, and the quantity $ne^2/m$, which is the Drude weight up to a constant. Importantly, these internal properties can be varied nearly independently, even within a single sample, for example, by changing the temperature or carrier concentration~\cite{Kukushkin1989}. At the same time, the frequency $\omega$ can be considered an extraneous parameter as it pertains to the external radiation. An additional intrinsic parameter of a disk is the radius. However, as a matter of convenience, we introduce the following dimensionless parameters
\begin{equation} \label{eq:Param}
	\widetilde{\gamma} = \frac{\gamma R}{c} = \frac{R}{c\tau}, \quad \widetilde{\Gamma} = \frac{2\pi n e^2R}{mc^2}, 
\end{equation}
where $c$ is the speed of light. These are the very parameters that determine the plasma resonances in the disk~\cite{Zagorodnev2021}. We refer to $\widetilde{\Gamma}$ as the retardation parameter. If it is small enough, for example, when the concentration is very low, then the standard quasistatic approximation is applicable in order to find resonant plasma frequency. In what follows, we normalize all frequencies by $c/R$ and all lengths by $R$, designating all dimensionless quantities with a tilde sign. For instance, the dimensionless frequency becomes $\widetilde\omega = \omega R/c$.

To find a response of the system, we must connect the induced field with the current. We derive this relation from Maxwell's equations, following the procedure described in detail in Ref.~\cite{Zagorodnev2021}. Given the cylindrical symmetry, the system can be characterized by the angular momentum $l$ and the radial number $n_r$. Therefore we can write Maxwell's equations in CGS units for the scalar and vector potentials in the Lorenz gauge using the cylindrical coordinates $\left(r,\theta,z\right)$. Since the radial and azimuthal components of the vector potential $\bm{A}$ are intermixed in this system, we apply a unitary transformation $S$ as follows:
\begin{equation} \label{eq:S-transform}
	S = \frac{1}{\sqrt{2}}
	\left(\begin{array}{cc}
		i & -i\\
		1 & 1
	\end{array}\right),
	\quad
	\bm{A}(r,z) = S \bm{A}_S(r,z),
\end{equation}
to diagonalize the system of equations. Note, that the $z$ component of the vector potential is zero. After that we apply the Hankel transform and solve the resulting equations with respect to the $z$ coordinate, taking into account boundary conditions on the metal plane $\bm{A}(\bm{r},-d)=\bm{A}_S(\bm{r},-d)=0$. Then, we express the transformed vector potential in terms of the transformed current, take the inverse Hankel transform, and do the inverse transformation $S$ to obtain the vector potential in the disk plane. Using it we find the desired (nonlocal) relation between the electric field induced in the disk plane $z=0$ and the current density
\begin{equation}\label{eq:InducedE}
	\bm{E}^{ind}(r)=i\frac{2\pi}{\omega}\left(\widehat{D}+\frac{\omega^2}{c^2}\right)\int\limits_{0}^{R}G_l(r,r')\bm{j}(r')r'dr',
\end{equation}
where the differential operator $\widehat{D}$ is given by
\begin{equation}\label{eq:operatorD}
	\widehat{D}=
	\left(\begin{array}{cc}
		\frac{\partial^2}{\partial r^2}+\frac{1}{r}\frac{\partial}{\partial r}-\frac{1}{r^2} & \frac{il}{r}\left(\frac{\partial}{\partial r}-\frac{1}{r}\right)\\
		\frac{il}{r}\left(\frac{\partial}{\partial r}+\frac{1}{r}\right) & -\frac{l^2}{r^2}
	\end{array}\right),
\end{equation}
and the kernel is
\begin{multline}\label{eq:kernel}
	G_l(r,r')=\int\limits_{0}^{\infty}\frac{pdp}{\beta}\left(1-e^{-2\beta d}\right)\cdot\\
	S\text{diag}\left[J_{l+1}(pr)J_{l+1}(pr'),J_{l-1}(pr)J_{l-1}(pr')\right]S^{-1}.
\end{multline}
Here $\text{diag}\left[\cdot,\cdot\right]$ denotes a $2\times2$ diagonal matrix, $J_l(z)$ is the Bessel function of the first kind of the $l$-th order, and $\beta=\sqrt{p^2-{\omega^2}/{c^2}}$, with the negative imaginary part in the branch of the square root since it corresponds to the electromagnetic waves outgoing from the disk.

The kernel $G_{l}\left(r,r'\right)$ has the same parity $(-1)^{l+1}$ as the Bessel functions of the order $l\pm 1$. Since the differential operator $\hat{D}$ does not alter the parity, the induced field and the current density can be chosen as odd and even functions for $l=0$ and $l=1$, respectively.

To calculate the disk response to the external electric field, we have to solve Eqs.~(\ref{eq:Current}) and (\ref{eq:InducedE}) with the boundary condition $j_r(R) = 0$. To find an approximate solution, we expand the unknown vector-function ${\bf j}(r)$ in a complete set of basis functions, taking into account the system parity, the boundary condition, and proper function behavior at the center of the disk. Then, we consecutively multiply it by the basis functions and integrate the system over the $r$ coordinate, reducing it to an infinite matrix equation for the expansion coefficients. After that, we truncate the matrix to retain only the dominant terms and solve it for the expansion coefficients. Consequently, we determine the desired current density that defines the system response. Finally, we calculate the absorption power
\begin{equation}\label{eq:AbsPower}
	P_{abs}=\int\limits_{0}^{R} \frac{1}{2}\text{Re}\left({\bf j^*}\cdot{{\bf E}^{tot}}\right)2\pi rdr=\frac{2\pi^2\widetilde{\gamma}}{c\widetilde{\Gamma}}\int\limits_{0}^{R}\left|{\bf j}\right|^2rdr,
\end{equation}
which provides us with information about the spectral position and width of the plasma resonances. The accuracy of this procedure can be assessed by the successive increase in the number of basis functions, however, we will see that a proper basis function describes all obtained features well.

\section{\label{sec:3}Eigenmodes in a fully screened limit}

Let us first address a simple case to provide some analytical results. Consider the eigenmodes (without external radiation) in a fully screened disk, where the distance $d$ is much less than all the other characteristic lengths. Indeed, for the low-frequency modes under consideration (with small $l$ and $n_r$), the next smallest length is the plasmon wavelength, which is about the size of the disk radius. Hence, assuming that $d \ll R$, we expand the term $1-e^{-2\beta d}$ in the integral kernel (\ref{eq:kernel}) as $2\beta d$. As will be shown below, this approximation excludes the radiation. It reduces the kernel to a delta function $\delta(r-r')/r'$, significantly simplifying the analysis because the relation between the induced field and the current becomes localized and is described by differential equations. It means that the charges within the disk interact with their images in the metal rather than with each other. Choosing the $S$ transformed basis, we can write Eq.~(\ref{eq:Current}) in the following form:
\begin{equation}\label{eq:ScreenedEq}
	-\frac{\mu^2}{R^2}\bm{j}_S(r) = \widehat{D}_S\bm{j}_S(r), \; \mu^2 = \widetilde{\omega}^2+\frac{i\widetilde{\omega}c}{4\pi \sigma\left(\widetilde{\omega}\right) \widetilde{d}},
\end{equation}
where $\widehat{D}_S = S^{-1}\widehat{D}S$ is the $S$-transformed differential operator (\ref{eq:operatorD}).

It can be shown that the vector-function $\left(J_{l+1}\left(\mu r/R \right),\ J_{l-1}\left(\mu r/R\right)\right)^T$ is the eigenvector of the operator $\widehat{D}_S$ and, therefore, is the solution to Eq.~(\ref{eq:ScreenedEq}). Here, the parameter $\mu$ is determined from the boundary condition at the edge of the disk $j_{S1}(R)=j_{S2}(R)$. Using the properties of Bessel functions and their derivatives, the boundary condition can be reduced to $J'_l(\mu)=0$ with the prime denoting the first derivative of the Bessel function by its argument. It means that $\mu$ is the $n_r$-th zero of the first derivative of the $l$-th order Bessel function \cite{Abramowitz1988}, which we denote as $\mu_{l,n_r}$.

From all of the above, we obtain the frequency of the eigenmodes in the following dimensionless form:
\begin{equation}\label{eq:omega_pl}
	\widetilde{\omega}_{l,n_r} = \sqrt{\frac{2{\mu^2_{l,n_r}}\widetilde{\Gamma}\widetilde{d}}{1+2\widetilde{\Gamma}\widetilde{d}}-\frac{\widetilde{\gamma}^2}{4\left(1+2\widetilde{\Gamma}\widetilde{d}\right)^2}}-i\frac{\widetilde{\gamma}}{2\left(1+2\widetilde{\Gamma}\widetilde{d}\right)}.
\end{equation}
where the imaginary part indicates the damping of the mode. In the quasistatic limit $\widetilde{\omega}_{l,n_r} \ll 1$, or $\widetilde{\Gamma}\widetilde{d} \ll 1$, the frequency of the eigenmodes was previously derived without dissipation in Ref.~\cite{Fetter1986}. It was also obtained in Ref.~\cite{Andreev2021}, including the electromagnetic retardation but still neglecting the damping. Now, we see that the damping of the resonances is determined by the collisions, however, it decreases with an increase in the role of retardation and the distance $d$. The reason is that in the quasistatic limit the plasmon energy exists only in the form of the kinetic energy of the charges and their potential interaction. As the parameter $\widetilde{\Gamma}\widetilde{d}$ increases, it also stores in electromagnetic energy surrounding the disk, i.e., the plasmon transforms into a plasmon-polariton. Therefore the damping decreases since the plasmon electromagnetic field is not subject to Ohmic losses. Nonetheless, it is clear that for any finite $d$, there are also radiative losses that will exceed those caused by the collisions at some critical value of $\widetilde{\Gamma}$. We consider it and estimate the critical parameters in the following sections.

\section{\label{sec:4}Axisymmetric mode ($l=0$)}

To excite the axisymmetric mode, a point dipole is placed at a distance $h$ over the center of the disk, with the dipole moment directed along the $z$-axis \cite{Duan2022}. Hence, it produces only the radial component of the electric field in the disk plane, which is given by 
\begin{equation*}
		E_r^{dip}(\widetilde{r},\widetilde{h})=\frac{d_0}{R^3} e^{i\widetilde{\omega}\widetilde{\rho}}\left(-\frac{3\widetilde{h}\widetilde{r}}{\widetilde{\rho}^5} + \frac{3i\widetilde{h}\widetilde{r}\widetilde{\omega}}{\widetilde{\rho}^4} + \frac{\widetilde{h}\widetilde{r}\widetilde{\omega}^2}{\widetilde{\rho}^3}\right),
\end{equation*}
where $\widetilde{\rho}=\sqrt{\widetilde{r}^2+\widetilde{h}^2}$, and $d_0$ is the absolute value of the electrical dipole moment. Note that the tilde sign marks the dimensionless parameters (for example, $\widetilde{r}=r/R$), as was mentioned in Sec.~\ref{sec:2}. The net external electric field includes the dipole radiation reflected by the metal, as follows
\begin{equation}\label{eq:l=0_ext}
	E_r^{ext}(\widetilde{r})=E_r^{dip}(\widetilde{r},\widetilde{h})+E_r^{dip}(\widetilde{r},-2\widetilde{d}-\widetilde{h}).
\end{equation}

To find the system response to the external excitation, we expand the current density in a set of functions. Since the external field in the disk plane is axisymmetric, we use the following expansion for the radial current
\begin{equation}\label{eq:l=0_functions}
	j_r(\widetilde{r})=\sum_{n=1}^N\alpha_n\widetilde{r}\left(1-\widetilde{r}^2\right)^n.
\end{equation}
It is merely an odd power Taylor series in which each basis function satisfies the boundary condition at the edge of the disk and behaves reasonably at the center of the disk. Next, we substitute the expansion (\ref{eq:l=0_functions}) into Eqs.~(\ref{eq:Current}),(\ref{eq:InducedE}) and consecutively multiply the obtained equation by the basis functions and integrate it over the radius $r$ using the weight function $r$.

We choose the polynomial basis primarily because its functions can be integrated analytically with respect to the coordinates $r$, $r'$ with the inner kernel of the integral operator $G_l(r,r')$. The remaining integrals over $p$ can be taken exactly for any frequency $\omega$ in the limits $d \rightarrow 0$ (as discussed in  Sec.~\ref{sec:3}) and $d \rightarrow \infty$ (as shown in Ref.~\cite{Zagorodnev2021}). They also can be found numerically but for small $\omega$ and any $d$ we can take them approximately, expanding them up to the third order in $\widetilde{\omega}$, which takes into account radiation by the disk. Indeed, it is beyond the standard quasistatic approximation mainly because of this radiation adjustment. It will be used to get the approximate estimates of the frequency and line width. The details of the expansion are provided in Appendix A.

After taking the integrals, we arrive at a system of $N$ equations for the coefficients $\alpha_n$. Solving the set of equations, we find the current density and then, according to Eq.~(\ref{eq:AbsPower}), the absorption power. We divide it by the power of the dipole radiation incident onto the disk to obtain the absorption coefficient $A=P_{abs}/P_{dip}$. Since the position and the line width of the plasmon resonance are internal characteristics of the electron system, they are almost independent on the distance to the dipole. Therefore, we place the disk in the far field region of the dipole for simplicity. Then the average radiation power incident on the disk is~\cite{Landau1975}
\begin{equation*}
	P_{dip} = \frac{1}{16}\frac{\omega^4 d_0^2}{c^3}\left(\frac{R}{h}\right)^4, \quad h \gg R.
\end{equation*}
The details of our calculations are given in Ref. \footnote{Wolfram Mathematica code to carry out the analysis is publicly available at https://github.com/danilrodionov/gated-disk.git}.

\begin{figure}
	\includegraphics[width=1\linewidth]{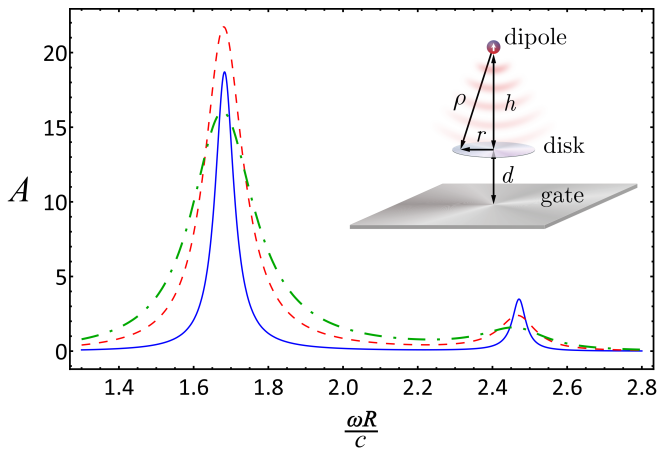}
	\caption{\label{fig:1} 
		(Color online) Frequency dependence of the absorption coefficient, $A$, of the gated conductive disk. The distance $d$ between the disk and the gate equals the disk radius $R$, the distance between the dipole and the disk is $h=100R$, i.e. the disk is in the far field region of the dipole radiation. The dipole moment is normal to the disk and the gate. It excites the axisymmetric plasma resonances. Clearly visible two resonances with radial numbers $n_r=1,2$. The retardation parameter, Eq.~(\ref{eq:Param}), is $\widetilde{\Gamma}=1$. Five terms of the current expansion series in Eq.~(\ref{eq:l=0_functions}) were used. Three curves correspond to the following electron collision rates: $\gamma R/c =0.02$ (blue solid), $0.1$ (red dashed), and $0.2$ (green dot-dashed). The inset (top right corner) schematically illustrates the excitation scheme.
	}
\end{figure}

As an example, Fig.~\ref{fig:1} shows the absorption spectra calculated for different collisional damping rates using the first five basis functions, at $\widetilde{\Gamma}=1$ and $d=R$. The data clearly indicate the first two radial resonances with $n_r = 1,2$. The absorption coefficient at the resonances is greater than unity since the disk collects light from the spatial region extending beyond its geometrical area. For instance, it is well known that for the dipole-like modes, the effective area (absorption cross section) of an arbitrarily small resonator can reach a value of the order of $\lambda^2$, where $\lambda$ is the wavelength of the incident radiation \cite{Mylnikov2022,Tretyakov2014}. Absorption reaches a maximum when the radiative losses are equal to the collisional ones. For different peaks, which correspond to different radial number, this happens at different values of the parameters.  We can see how the absorption is maximized with increasing collisional scattering by comparing the data curves at the main resonance ($n_r = 1$).

\begin{figure}
	\includegraphics[width=1\linewidth]{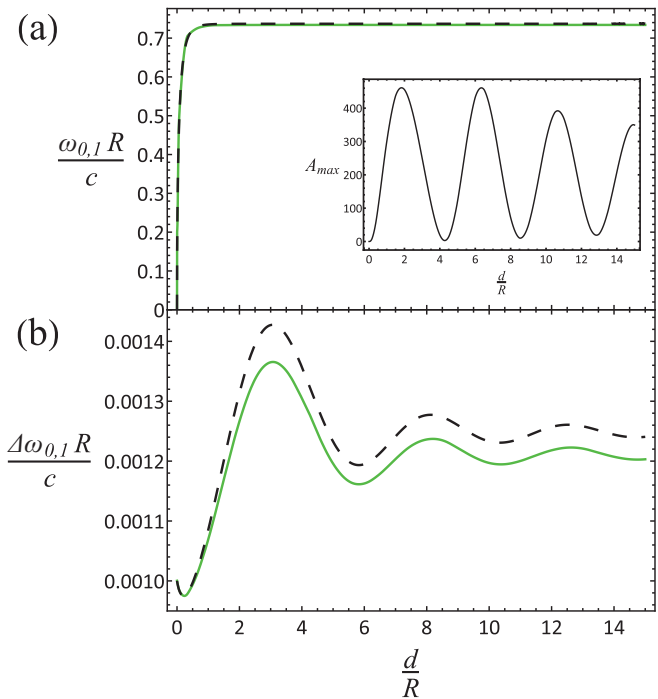}
	\caption{\label{fig:2}  
		(Color online) Dependence of the main axisymmetric resonance frequency $\omega_{0,1}$ (a) and its line width $\Delta\omega_{0,1}$ (b) on the distance $d$ between the gate and the disk of radius $R$. The data are obtained for the dimensionless retardation parameter $\widetilde{\Gamma} = 0.16$ (nearly quasistatic), the dimensionless electron collisional damping rate $\gamma R/c = 0.001$, and $h=100R$. The solid and dashed-line curves compare, respectively, the numerical calculations with five basis functions from the set (\ref{eq:l=0_functions}) and the analytical result based on one basis function (see Appendix A). The inset represents the dependence of the absorption coefficient at the resonance frequency on the separation distance $d$.}
\end{figure}

Below in this section, we focus only on the main resonance with $n_r=1$. From the dependence of the absorption coefficient on the frequency, we find the resonant frequency $\widetilde{\omega}_{0,1}$, taken at the maximum of the first absorption peak, and the line width $\Delta\widetilde{\omega}_{0,1}$, taken as the full width at half maximum. In Fig.~\ref{fig:2}, it is shown how characteristics of the resonance depend on the distance between the disk and the gate. We can see that the frequency monotonically transitions from the gated ($d \ll R$) to the ungated ($d \gg R$) regime, where it becomes completely independent of $d$. The line width initially indicates a slight decrease for the small values of $d$, which is in accordance with Eq.~(\ref{eq:omega_pl}). Then, it starts to increase as the radiative broadening becomes dominant --- for the given parameters (Fig.~\ref{fig:2}), the transition occurs at $d \sim R$. However, the lower the collisional scattering rate, the smaller the critical distance. At larger values of $d$, the line width is governed mainly by the radiative losses, exhibiting oscillatory behavior due to the constructive or destructive interference between the fields from the disk and the metal. We note that for $d>R$, the resonant frequency is almost constant, whereas the line width oscillations are still noticeable, even when the distance is several times larger than the disk radius. Finally, as $d\rightarrow\infty$, the line width reaches the asymptotic value corresponding to the ungated disk.

To reinforce the results described above, we also provide analytical approximations considering only the first basis function from the set (\ref{eq:l=0_functions}). As shown in Fig.~\ref{fig:2}, these estimates (plotted in the dashed lines) closely follow the data obtained using five basis functions. In this case, the absorption coefficient is proportional to $|\alpha_1|^2$ and we find the analytical expressions for the resonant frequency and line width by minimizing its denominator (for the derivation details, see Appendix A). In the limit of the small separation distance, $d \ll R$, we arrive at the following approximations for the dimensionless frequency and line width
\begin{equation} \label{eq:FreqSmall}
	\widetilde{\omega}_{0,1} \approx 4\sqrt{2\widetilde{\Gamma}\widetilde{d}}\left( 1-\widetilde{d}\widetilde{\Gamma}+\frac{3}{2}\widetilde{\Gamma}^2\widetilde{d}^2 + ... \right),
\end{equation} 
and the dimensionless line width
\begin{equation} \label{eq:WidthSmall}
	\Delta\widetilde{\omega}_{0,1} \approx \widetilde{\gamma} \left( 1-2\widetilde{\Gamma}\widetilde{d}+4\widetilde{\Gamma}^2\widetilde{d}^2 + ... \right) + 156 \widetilde{d}^5 \widetilde{\Gamma}^4 +... \quad .
\end{equation}
These results are in excellent agreement with Eq.~(\ref{eq:omega_pl}). The only difference between the frequencies is the proportionality coefficient of $4$ instead of $\mu_{0,1}\approx 3.83$ in Eq.~(\ref{eq:omega_pl}). The part of the line width proportional to $\widetilde{\gamma}$ also matches the expansion of the imaginary part in (\ref{eq:omega_pl}). Indeed, the only difference between the line widths is in the last term of Eq.~(\ref{eq:WidthSmall}) that accounts for the plasmon radiation. Hence, the standard quasistatic approximation is valid for the description of the plasmon damping in a gated disk if the 'fully screened' damping $\widetilde{\gamma}/(1+2\widetilde{\Gamma}\widetilde{d})$ is less than the last term in Eq.~(\ref{eq:WidthSmall}). In fact, this term is the octupole radiation of the system. It has been shown that the axisymmetric mode in a bare disk has a zero dipole moment ($\propto\widetilde{\Gamma}^2$) and a non-zero quadrupole moment ($\propto\widetilde{\Gamma}^3$) \cite{Zagorodnev2019,Zagorodnev2021}. However, the nearby gate suppresses the quadrupole radiation, making the octupole radiation ($\propto\widetilde{\Gamma}^4$) dominant.

Equation (\ref{eq:WidthSmall}) also clearly demonstrates that the total absorption line width is not merely the sum of the collisional ($\gamma$) and radiative broadenings in contrast to the common assumption \cite{Zoric2011,Andreev2014}. The thing is that the plasmon damping depends on the retardation parameter, $\widetilde{\Gamma}$, even when the plasmon does not radiate electromagnetic waves, as was shown in Eq. (\ref{eq:omega_pl}) for the fully screened limit. Therefore there is no pure ``collisional'' broadening when the retardation is taken into account and the part of the broadening proportional to the collisional decay rate depends on the retardation parameter as well, while the radiative broadening is also determined by the retardation parameter. That is why the line width is a nontrivial intermixture of the collisional and radiative broadenings.

In the case of $d \gg R$, the resonance frequency and line width can be approximated as
\begin{equation} \label{eq:FreqHuge}
	\widetilde{\omega}_{0,1} \approx 1.87\sqrt{\widetilde{\Gamma}}\left(1-0.086\widetilde{\Gamma}-0.011\widetilde{\Gamma}^2\right)+... ,
\end{equation}
\begin{equation} \label{eq:WidthHuge}
	\Delta\widetilde{\omega}_{0,1} \approx \Delta\widetilde{\omega}_{0,1}^{\infty} + 1.02\widetilde{\Gamma}^3\frac{\sin 2\widetilde{\omega}_{0,1}\widetilde{d}}{\left(2\widetilde{\omega}_{0,1}\widetilde{d}\right)^2}+...\;.
\end{equation}
Here, $\Delta\widetilde{\omega}_{0,1}^{\infty}$ is the line width for a bare disk, which is independent of $d$. In the case under consideration, $\widetilde{\gamma} \ll \widetilde{\Gamma} \ll 1$, it is also proportional to $0.14\widetilde{\Gamma}^3$ \cite{Zagorodnev2021}. The second term in Eq.(\ref{eq:WidthHuge}) is determined by the quadrupole radiation, the amplitude of which depends on the interference of electromagnetic fields created by the disk and the metal charges. For large values of $d$, it has the form of damped oscillations with the period defined by the parameter $\omega_{0,1}d/c$, which is almost constant at $d > R$. In a 'clean' limit, $\widetilde{\gamma}=0$, the magnitude of the first maximum is approximately $1.7$ times greater than $\Delta\widetilde{\omega}_{0,1}^{\infty}$.

\section{\label{sec:5}Fundamental mode ($l=1$)}

In this section, we apply the approach developed above to the fundamental resonance, with $n_r=1$ and $l=1$. As an excitation, we choose a circularly polarized plane wave with the electrical field expressed in cylindrical coordinates as follows:
\begin{equation}\label{eq:l=1_ext}
	\bm{E}^{ext}(\widetilde{r})=E_0(1,\,i)^T\sin\left(\widetilde{\omega}\widetilde{d}\right).
\end{equation}
We consider the $S$-transformed basis, where the integral kernel (\ref{eq:kernel}) is purely diagonal. We use polynomials to expand the transformed current $\bm{j}_S(r)$ as they can be analytically integrated with Bessel functions in the integral kernel. Thus we choose the following series for the transformed current density
\begin{equation}\label{eq:l=1_functions}
	\bm{j}_S(\widetilde{r})=\alpha_1
	\left(
	\begin{array}{c}
		\widetilde{r}^2\\
		1
	\end{array}
	\right)+\sum_{n=1}^{N}
	\left(
	\begin{array}{c}
		\alpha_{2n}\widetilde{r}^2\\
		\alpha_{2n+1}
	\end{array}
	\right)\left(1-\widetilde{r}^2\right)^n,
\end{equation}
where $\alpha_n$ are the expansion coefficients to be calculated. These polynomials satisfy the boundary condition and the parity of the kernel (\ref{eq:kernel}). Hence, applying the method described in the previous section, we first construct a system of linear equations for the unknown coefficients. Next, we consecutively multiply the $S$-transformed equation for the current by every basis function from the set (\ref{eq:l=1_functions}), and integrate it over $r$ with the weight function $r$. Then, we solve the linear system and obtain the current density needed to calculate the resonant frequency, and the line width.

\begin{figure}
	\includegraphics[width=1\linewidth]{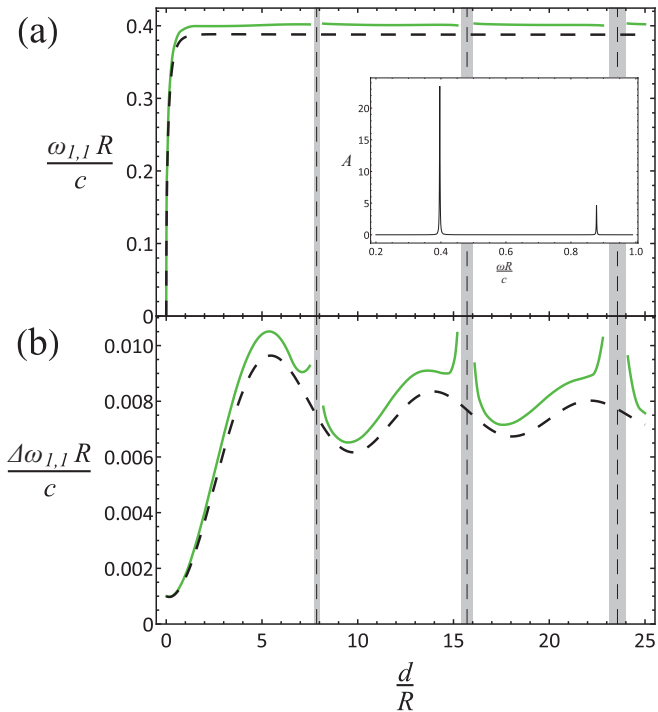}
	\caption{\label{fig:3} 
		(Color online) Dependence of the main fundamental resonance frequency $\omega_{1,1}$ (a) and its line width $\Delta\omega_{1,1}$ (b) on the distance $d$ between the gate and the disk of radius $R$. The data are obtained for the dimensionless retardation parameter $\widetilde{\Gamma} = 0.16$ and the dimensionless electron collisional damping rate $\gamma R/c = 0.001$. The solid (green) and dashed (black) curves are, respectively, the numerical calculations with eleven basis functions from the set (\ref{eq:l=1_functions}) and the analytical results based on a single basis function. The vertical dashed lines mark the frequencies $n\pi c/d$, where $n$ is a natural number, in which the net external field in the disk plane is zero due to destructive interference from the incident plane wave and its reflection.The grey shaded areas are the regions of substantial resonance deformation. The inset represents the dependence of the absorption coefficient on the frequency at $d=R$.}
\end{figure}

In Fig.~\ref{fig:3}, we plot the frequency and line width dependence on the gate separation distance for the fundamental resonance. An example of the absorption spectrum is shown in the inset. The numerical calculations (solid curves) are based on eleven basis functions (i.e., $N=5$) and the external field in Eq.~(\ref{eq:l=1_ext}). We use the retardation parameter $\widetilde{\Gamma} = 0.16$, and the collisional decay rate $\widetilde{\gamma} = 0.001$. For comparison, we also plot the analytical estimates (dashed curves) according to Eqs.~(\ref{ap:l=1_pos}) and (\ref{ap:l=1_wid}) provided in the Appendix B, obtained using a single basis function (i.e., $N=0$) and, for simplicity, do not take into account the sine function in the external field (\ref{eq:l=1_ext}). Similar to the previous section, we use $\widetilde{\Gamma} \ll 1$ and consider only the denominator of the absorption coefficient.

The analytical and numerical results show good agreement, with the exception of the grey shaded regions where the resonance line is strongly affected by the vanishing of the total external radiation due to the sine function in Eq.~(\ref{eq:l=1_ext}). Indeed, the absorption line shape near the plasma resonances can be approximated as the Lorentzian function, arising from the resonant denominators in the expansion coefficients $\alpha_n$, modulated by the term $\sin^2(\omega d/c)$, since the absorption is proportional to $\left|\bm{E}^{ext}\right|^2$. When a resonance frequency of the Lorentzian approaches to the zeros of sine, i.e., the frequencies $n\pi c/d$, where $n$ is a natural number, the absorption line first becomes asymmetric, what does not noticeably affect the position of the resonance, but significantly increases the line width. This increase is clearly visible in Fig.~\ref{fig:3}(b) near the grey areas. The centers of the grey area (dashed vertical lines) correspond to the zeros of the sine function, i.e., the distances $d_n=n\pi c/\omega_{0,1}$. The boundaries of the regions are established by the inequality $|\omega_{0,1}-n\pi c/d|<\Delta\omega_{0,1}/2$, where the line width $\Delta\omega_{0,1}$ is taken at the distance $d_n$. It means that inside the grey regions the Lorentzian is so strongly affected by the sine function that the resonance line is divided into two peaks. Since this happens due to the peculiarity of the excitation scheme and is not related to the characteristics of the plasma waves, we do not show any numerical data in the grey shaded area.

\begin{figure}
	\includegraphics[width=1\linewidth]{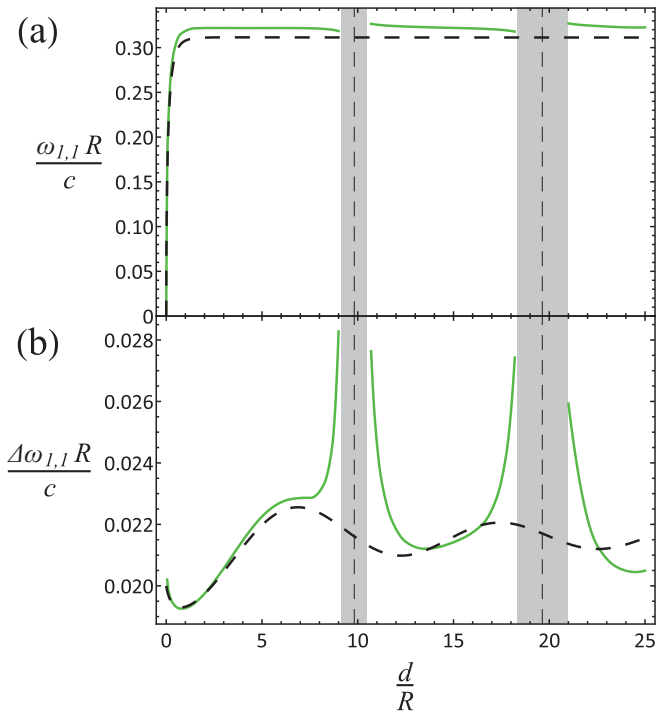}
	\caption{\label{fig:4} 
		(Color online) Dependence of the main fundamental resonance frequency $\omega_{1,1}$ (a) and its line width $\Delta\omega_{1,1}$ (b) on the distance $d$ between the metal gate and the disk. The data are obtained for the dimensionless retardation parameter $\widetilde{\Gamma} = 0.1$ and the dimensionless electron collisional damping rate $\gamma R/c= 0.02$. The solid (green) curves represent the numerical calculations with eleven basis functions from the set (\ref{eq:l=1_functions}). The dashed (black) curves correspond to the analytical result based on one basis function.}
\end{figure}

In real systems, however, the effect can be reduced by the nonideal conductivity of the metal, the absorption in the dielectric medium between the disk and the gate, or the divergence of the external radiation. For example, the discussed above dipole excitation of axisymmetric mode, which naturally contains the divergence, completely removes it. At the same time, a wider resonance degrades the oscillatory pattern, as illustrated by the data for $\widetilde{\gamma}=0.02$ in Fig.~\ref{fig:4}. In this case, the line width oscillations are quite distorted, though the frequency behavior is preserved.

Now, let us discuss the two simple analytical limits. In the first limit of $d\ll R$, the resonance frequency and line width are given as
\begin{equation}\label{eq:l=1_pos_sd}
	\widetilde{\omega}_{1,1} \approx 1.73\sqrt{2\widetilde{\Gamma}\widetilde{d}}\left(1-\widetilde{\Gamma}\widetilde{d}+\frac{3}{2}\widetilde{\Gamma}^2\widetilde{d}^2+...\right),
\end{equation}
and
\begin{equation}\label{eq:l=1_wid_sd}
	\Delta\widetilde{\omega}_{1,1} \approx \widetilde{\gamma}\left(1-2\widetilde{\Gamma}\widetilde{d}+4\widetilde{\Gamma}^2\widetilde{d}^2+...\right)+7.2\widetilde{\Gamma}^3\widetilde{d}^4+... \;. 
\end{equation}
These expressions match well the expansion of (\ref{eq:omega_pl}) at $d\ll R$. The only noticeable difference in the frequency is the proportionality coefficient of $1.73$ as opposed to the value $\mu_{1,1} \approx 1.84$. The part of the line width proportional to $\widetilde{\gamma}$ is also consistent with the expansion of the imaginary frequency component in (\ref{eq:omega_pl}) at small $\widetilde{\Gamma}\widetilde{d}$. The extra $\widetilde{\Gamma}^3$ term in (\ref{eq:l=1_wid_sd}) corresponds to the quadrupole radiation, whereas the dipole radiation component proportional to $\widetilde{\Gamma}^2$ is suppressed by the nearby image charges.

In the second limit of $d\gg R$, when the gate is far away from the disk, the resonance frequency and line width can be approximated as follows:
\begin{equation}\label{eq:l=1_pos_id}
	\widetilde{\omega}_{1,1} \approx 1.01\sqrt{\widetilde{\Gamma}}\left(1-0.25\widetilde{\Gamma}+...\right),
\end{equation}
\begin{equation}\label{eq:l=1_width_id}
	\Delta\widetilde{\omega}_{1,1} \approx \Delta\widetilde{\omega}_{1,1}^\infty-0.19\widetilde{\Gamma}^2\frac{\sin 2\widetilde{\omega}_{1,1}\widetilde{d}}{\widetilde{\omega}_{1,1}\widetilde{d}}+...\;.
\end{equation}
Here, $\Delta\widetilde{\omega}_{1,1}^\infty$ is the line width in the absence of the metal. It can be estimated by taking the limit of $d\rightarrow\infty$ in Eq.~(\ref{ap:l=1_wid}). The second term in Eq.~(\ref{eq:l=1_width_id}) is the radiative damping determined by the dipole radiation. It is an oscillatory function of the parameter $\omega_{1,1}d/c$, which can be explained by the interference between the radiation outgoing from the disk and its reflection from the gate. In the clean limit, $\widetilde{\gamma}=0$, the magnitude of the first maximum is approximately $1.35$ times greater than $\Delta\widetilde{\omega}_{1,1}^{\infty}$. We also note that the frequency in (\ref{eq:l=1_pos_id}) is almost independent of $d$, approaching the asymptotic value of the ungated limit.

\section{\label{sec:6}Discussion and conclusion}
To put the presented results into a practical perspective, let us provide some estimates of achievable parameters for a disk of the 2D electron gas. Thus, for GaAs/AlGaAs quantum wells with the typical 2D electron concentration $n \sim 2\cdot 10^{11}$ cm$^{-2}$, the effective mass $0.066m_e$, and the disk radius $R \sim 0.3$ mm, we obtain the retardation parameter $\widetilde{\Gamma} \sim 0.16$. In high-mobility samples \cite{Gusikhin2018}, the electron relaxation time can reach the values of up to $10^{-10}$ s at the liquid-helium temperatures, yielding the dimensionless relaxation rate $\widetilde{\gamma} \sim 0.01$, although it can be reduced by order of magnitude in ultrahigh quality samples at lower temperatures \cite{Chung2021}. Similar estimates can be made for the quantum wells based on ZnO or InAs  \cite{Falson2016,Tschirky2017}. Therefore the parameter values considered in our work are feasible for real structures, and the intrinsic line width of the main resonances is governed mainly by the radiation when the gate is sufficiently far away from the 2D electron system.

In summary, we have studied numerically and analytically the fundamental and axisymmetric plasma resonances in a conductive disk in the presence of an ideal metal gate. Considering high-mobility samples, we find that, unlike the frequencies, the damping of the modes exhibits non-monotonic dependence on the distance between the gate and the disk. When the distance increases, the damping is initially reduced due to the growing role of electromagnetic retardation and the storage of plasmon energy in an ambient medium. Then, it becomes stronger as the radiative contribution begins to dominate. After that, the damping oscillates because of the constructive and destructive interference between the radiation from the disk and its image, finally reaching the asymptotic value of the ungated disk.

\begin{acknowledgments}
	This work was supported by the Russian Science Foundation (project No. 21-12-00287). We are grateful to V. M. Muravev, A.A. Zabolotnykh, and V. A. Volkov for valuable discussions.
\end{acknowledgments}

\appendix
\section{Approximations for axisymmetric mode}

Here we give the details of approximate calculations of the frequency and line width based on one basis function in the set (\ref{eq:l=0_functions}). Applying the procedure described in Sec.~\ref{sec:4} (the Galerkin method) we get the following expression for the coefficient:
\begin{equation}\label{ap:l=0_alpha}
	\alpha_1=\frac{12\widetilde{\Gamma}c\int\limits_0^1\widetilde{r} d\widetilde{r} E_r^{ext}(\widetilde{r})\widetilde{r}(1-\widetilde{r}^2)}{\pi\left[\widetilde{\omega}+i\widetilde{\gamma}-\frac{24\widetilde{\Gamma}}{\widetilde{\omega}}f(\widetilde{\omega})\right]}.
\end{equation}
Here function $f(\widetilde{\omega})$ is
\begin{equation*}
	f(\widetilde{\omega}) = \int\limits_{0}^{\infty}dp\widetilde{\beta}p\left(1-e^{-2\widetilde{\beta}d}\right)F^2(p),
\end{equation*}
where
\begin{equation*}
	F(p)=\int\limits_{0}^1\widetilde{r}d\widetilde{r}J_1(p\widetilde{r})\widetilde{r}\left(1-\widetilde{r}^2\right)=\frac{2J_3(p)}{p^2}
\end{equation*}
is the Hankel tranform of the current density, and $\widetilde{\beta} = \sqrt{p^2-\widetilde{\omega}^2}$. Expanding the function $f(\widetilde{\omega})$ in the Taylor series, we get
\begin{equation*}
	f(\widetilde{\omega}) \approx \frac{16 a_0}{35\pi}-\frac{64 a_2}{2\,835\pi}\widetilde{\omega}^2-\frac{256 a_4}{155\,925\pi}\widetilde{\omega}^4-i\frac{a_5}{4\,320}\widetilde{\omega}^5,
\end{equation*}
where coefficients are
\begin{widetext}
	\begin{multline*}
		a_0 \approx 1+9.16 \widetilde{d}-\widetilde{d}\left(1-27.7\widetilde{d}^2-33.8\widetilde{d}^4-14.2\widetilde{d}^6\right)E\left(-\frac{1}{\widetilde{d}^2}\right)-\widetilde{d}\left(17+43.7\widetilde{d}^2+40.9\widetilde{d}^4+14.2\widetilde{d}^6\right)K\left(-\frac{1}{\widetilde{d}^2}\right),
	\end{multline*}
	\begin{multline*}
		a_2 \approx 1+11.6 \widetilde{d}+247\widetilde{d}^3-\widetilde{d}\left(1+60.5\widetilde{d}^2-364\widetilde{d}^4-268\widetilde{d}^6+80\widetilde{d}^8\right)E\left(-\frac{1}{\widetilde{d}^2}\right)+\nonumber\\
		\widetilde{d}\left(0.5-265\widetilde{d}^2-493\widetilde{d}^4-308\widetilde{d}^6-80\widetilde{d}^8\right)K\left(-\frac{1}{\widetilde{d}^2}\right),
	\end{multline*}
	\begin{multline*}
		a_4 \approx 1-106 \widetilde{d}^3-1021\widetilde{d}^5-\widetilde{d}\left(1-20.9\widetilde{d}^2-343\widetilde{d}^4+993\widetilde{d}^6+520\widetilde{d}^8+120\widetilde{d}^{10}\right)E\left(-\frac{1}{\widetilde{d}^2}\right)+\nonumber\\
		\widetilde{d}\left(0.5-10.4\widetilde{d}^2+775\widetilde{d}^4+125\widetilde{d}^6+580\widetilde{d}^8+120\widetilde{d}^{10}\right)K\left(-\frac{1}{\widetilde{d}^2}\right),
	\end{multline*}
	\begin{multline}\label{ap:l=0_coef}
		a_5 \approx \frac{1+3.75\widetilde{d}\widetilde{\omega}\left(\widetilde{d}^2\widetilde{\omega}^2-3\right)\cos\left(2\widetilde{d}\widetilde{\omega}\right)+9.38\left(0.6-\widetilde{d}^2\widetilde{\omega}^2\right)\sin\left(2\widetilde{d}\widetilde{\omega}\right)}{\widetilde{d}^5\widetilde{\omega}^5}+\\
		1.88i\frac{3+\widetilde{d}^2\widetilde{\omega}^2+5\left(\widetilde{d}^2\widetilde{\omega}^2-0.6\right)\cos\left(2\widetilde{d}\widetilde{\omega}\right)+2\widetilde{d}\widetilde{\omega}\left(\widetilde{d}^2\widetilde{\omega}^2-3\right)\sin\left(2\widetilde{d}\widetilde{\omega}\right)}{\widetilde{d}^5\widetilde{\omega}^5}.
	\end{multline}
\end{widetext}
Here the functions $K(k)$ and $E(k)$ are the complete elliptic integrals of the first and second kind respectively. All coefficients above are equal to unity in the limit of $\widetilde{d}\rightarrow \infty$ (in the absence of a metal). The coefficients $a_0$, $a_2$, and $a_4$ are monotonically rising functions of the distance $\widetilde{d}$, whereas the coefficient $a_5$ has oscillating dependence on the distance $\widetilde{d}$.

The absorption power is proportional to the square of the density current, i.e., $|\alpha_1|^2$. The numerator of the coefficient is determined by the external field, therefore it depends on the frequency weakly and smoothly, while the denominator significantly depends on the frequency and determines the resonance. Therefore, considering only denominator of the $|\alpha_1|^2$ and high quality resonances ($\widetilde{\gamma}\ll\widetilde{\Gamma}$), we approximate the absorption resonance by the Lorentz contour and arrive to the following expressions for the resonance position
\begin{multline}\label{ap:l=0_pos}
	\widetilde{\omega}_{0,1} \approx 1.87\sqrt{a_0\widetilde{\Gamma}}\Big(1-0.086 a_2\widetilde{\Gamma}-\\
	-0.011\left[2.04 a_0 a_4-1.04 a_2^2\right]\widetilde{\Gamma}^2\Big) ,
\end{multline}
and the line width
\begin{multline}\label{ap:l=0_wid}
	\Delta\widetilde{\omega}_{0,1} \approx \widetilde{\gamma}\Big[1-0.172 a_2\widetilde{\Gamma}-0.058\left(1.51 a_0 a_4-0.51a_2^2\right){\Gamma}^2+\\
	0.032 a_2\left(1.11a_0 a_4-0.11a_2^2\right)\widetilde{\Gamma}^3\Big]
	+0.068 a_0^2 \text{Re}\left(a_5\right)\widetilde{\Gamma}^3 .
\end{multline}
In these equations the coefficient $a_5$ and its derivative $a_5'$ by the parameter $\widetilde{\omega}\,\widetilde{d}$ should be calculated at the resonance frequency (\ref{ap:l=0_pos}). We use this approximate expressions to plot the dashed curve in Fig.~\ref{fig:2}. Expanding it for $\widetilde{d} \ll 1$ and $\widetilde{d} \rightarrow \infty$ we obtain Eqs.~(\ref{eq:FreqSmall})-(\ref{eq:WidthHuge}).

\section{Approximations for fundamental mode}

Here we get the analytical approximation of the frequency and line width based on the first basis function from the set (\ref{eq:l=1_functions}). Applying the solution method described in Sec.~\ref{sec:5}, we obtain the following expression for the first expansion coefficient
\begin{equation}\label{eq:l=1_alpha}
	\alpha_1=\frac{-3\widetilde{\Gamma}c E_0\sin\widetilde{\omega}\widetilde{d}}{8\pi\left[\widetilde{\omega}+i \widetilde{\gamma}-\frac{3}{4}\frac{\widetilde{\Gamma}}{\widetilde{\omega}}f(\widetilde{\omega})\right]}.
\end{equation}
The function $f(\widetilde{\omega})$ is given by
\begin{multline*}
	f(\widetilde{\omega})=\int\limits_{0}^{\infty}\frac{p}{\widetilde{\beta}}\left(1-e^{-2\widetilde{\beta}\widetilde{d}}\right)\Big[p^2\left(F_1(p)+F_2(p)\right)^2-\\
	2\widetilde{\omega}^2\left(F_1^2(p)+F_2(p)^2\right)\Big],
\end{multline*}
where
\begin{eqnarray}
	F_1(p)=\int\limits_{0}^{1}\widetilde{r}d\widetilde{r}J_2(p\widetilde{r})\widetilde{r}^2=\frac{J_3(p)}{p},\nonumber\\
	F_2(p)=\int\limits_{0}^{1}\widetilde{r}d\widetilde{r}J_0(p\widetilde{r})=\frac{J_1(p)}{p}\nonumber
\end{eqnarray}
are the Hankel transform of the components of the transformed current density. The expansion $f(\widetilde{\omega})$ at the small frequency $\widetilde{\omega}\ll 1$ is
\begin{equation*}\label{ap:l=1_f}
	f(\widetilde{\omega}) \approx \frac{64a_0}{15\pi}-\frac{656 a_2}{315\pi}\widetilde{\omega}^2-i\frac{a_3}{3}\widetilde{\omega}^3,
\end{equation*}
where coefficients are
\begin{widetext}
	\begin{equation*}
		a_0\approx1+5.89\widetilde{d}-\widetilde{d}\left(1-9\widetilde{d}^2-4\widetilde{d}^4\right)E\left(-\frac{1}{\widetilde{d}^2}\right)-\widetilde{d}\left(7+11\widetilde{d}^2+4\widetilde{d}^4\right)K\left(-\frac{1}{\widetilde{d}^2}\right),
	\end{equation*}
	\begin{multline*}
		a_2\approx1+4.02\widetilde{d}+8.05\widetilde{d}^3-\widetilde{d}\left(1-2.17\widetilde{d}^2-11.3\widetilde{d}^4-3.9\widetilde{d}^6\right)E\left(-\frac{1}{\widetilde{d}^2}\right) - \widetilde{d}\left(3.34+12.7\widetilde{d}^2+13.3\widetilde{d}^4+3.9\right)K\left(-\frac{1}{\widetilde{d}^2}\right),
	\end{multline*}
	\begin{multline}\label{ap:l=1_coef}
		a_3\approx\frac{\widetilde{\omega}^3\widetilde{d}^3-0.375\widetilde{\omega}\widetilde{d}\cos\left(2\widetilde{\omega}\widetilde{d}\right)+0.188\left(1-4\widetilde{\omega}^2\widetilde{d}^2\right)\sin\left(2\widetilde{\omega}\widetilde{d}\right)}{\widetilde{\omega}^3\widetilde{d}^3}+\\
		0.188i\frac{1-2\widetilde{\omega}^2\widetilde{d}^2-\left(1-4\widetilde{\omega}^2\widetilde{d}^2\right)\cos\left(2\widetilde{\omega}\widetilde{d}\right)-2\sin\left(2\widetilde{\omega}\widetilde{d}\right)}{\widetilde{\omega}^3\widetilde{d}^3}.
	\end{multline}
\end{widetext}

The coefficients (\ref{ap:l=1_coef}) in the absence of a metal ($\widetilde{d}\rightarrow\infty$) approach to unity. The coefficients $a_0$, $a_2$ are monotonically rising functions of the gate distance. The coefficient $a_3$ is an oscillating function of the parameter $\widetilde{d}$.

According to Eq.~(\ref{eq:AbsPower}), the absorption power is proportional to $|\alpha_1|^2$, therefore, the absorption resonance is determined by the denominator of the first coefficient. Considering high enough quality resonances, we approximate it by the Lorentz contour and extract the resonance position
\begin{equation}\label{ap:l=1_pos}
	\widetilde{\omega}_{1,1} \approx 1.01\sqrt{a_0\widetilde{\Gamma}}\left(1-0.25a_2\widetilde{\Gamma}\right),
\end{equation}
and the line width
\begin{multline}\label{ap:l=1_wid}
	\Delta\widetilde{\omega}_{1,1} \approx \widetilde{\gamma}\Big(1-0.5a_2\widetilde{\Gamma}+0.15a_2^2\widetilde{\Gamma}^2 \Big)+0.25a_0\text{Re}\left(a_3\right)\widetilde{\Gamma}^2
\end{multline}
at small retardation, $\widetilde{\Gamma}\ll 1$.
Here the coefficient $a_3$ and its derivative $a_3'$ by $\widetilde{\omega}\widetilde{d}$ are evaluated at the resonance frequency (\ref{ap:l=1_pos}). We use this approximate expressions to plot the dashed curve in Figs.~\ref{fig:3} and \ref{fig:4} and to derive
Eqs.~(\ref{eq:l=1_pos_sd})--(\ref{eq:l=1_width_id}) by expanding these expressions for $\widetilde{d} \ll 1$ and $\widetilde{d} \rightarrow \infty$.

\bibliography{main}

\begin{thebibliography}{40}%
\makeatletter
\providecommand \@ifxundefined [1]{%
 \@ifx{#1\undefined}
}%
\providecommand \@ifnum [1]{%
 \ifnum #1\expandafter \@firstoftwo
 \else \expandafter \@secondoftwo
 \fi
}%
\providecommand \@ifx [1]{%
 \ifx #1\expandafter \@firstoftwo
 \else \expandafter \@secondoftwo
 \fi
}%
\providecommand \natexlab [1]{#1}%
\providecommand \enquote  [1]{``#1''}%
\providecommand \bibnamefont  [1]{#1}%
\providecommand \bibfnamefont [1]{#1}%
\providecommand \citenamefont [1]{#1}%
\providecommand \href@noop [0]{\@secondoftwo}%
\providecommand \href [0]{\begingroup \@sanitize@url \@href}%
\providecommand \@href[1]{\@@startlink{#1}\@@href}%
\providecommand \@@href[1]{\endgroup#1\@@endlink}%
\providecommand \@sanitize@url [0]{\catcode `\\12\catcode `\$12\catcode
  `\&12\catcode `\#12\catcode `\^12\catcode `\_12\catcode `\%12\relax}%
\providecommand \@@startlink[1]{}%
\providecommand \@@endlink[0]{}%
\providecommand \url  [0]{\begingroup\@sanitize@url \@url }%
\providecommand \@url [1]{\endgroup\@href {#1}{\urlprefix }}%
\providecommand \urlprefix  [0]{URL }%
\providecommand \Eprint [0]{\href }%
\providecommand \doibase [0]{https://doi.org/}%
\providecommand \selectlanguage [0]{\@gobble}%
\providecommand \bibinfo  [0]{\@secondoftwo}%
\providecommand \bibfield  [0]{\@secondoftwo}%
\providecommand \translation [1]{[#1]}%
\providecommand \BibitemOpen [0]{}%
\providecommand \bibitemStop [0]{}%
\providecommand \bibitemNoStop [0]{.\EOS\space}%
\providecommand \EOS [0]{\spacefactor3000\relax}%
\providecommand \BibitemShut  [1]{\csname bibitem#1\endcsname}%
\let\auto@bib@innerbib\@empty
\bibitem [{\citenamefont {Pines}\ and\ \citenamefont {Bohm}(1952)}]{Pines1952}%
  \BibitemOpen
  \bibfield  {author} {\bibinfo {author} {\bibfnamefont {D.}~\bibnamefont
  {Pines}}\ and\ \bibinfo {author} {\bibfnamefont {D.}~\bibnamefont {Bohm}},\
  }\bibfield  {title} {\bibinfo {title} {A collective description of electron
  interactions: $\mathrm{II.}$ $\mathrm{C}$ollective vs individual particle
  aspects of the interactions},\ }\href
  {https://doi.org/10.1103/PhysRevLett.18.546} {\bibfield  {journal} {\bibinfo
  {journal} {Phys. Rev.}\ }\textbf {\bibinfo {volume} {85}},\ \bibinfo {pages}
  {338} (\bibinfo {year} {1952})}\BibitemShut {NoStop}%
\bibitem [{\citenamefont {Stern}(1967)}]{Stern1967}%
  \BibitemOpen
  \bibfield  {author} {\bibinfo {author} {\bibfnamefont {F.}~\bibnamefont
  {Stern}},\ }\bibfield  {title} {\bibinfo {title} {Polarizability of a
  two-dimensional electron gas},\ }\href
  {https://doi.org/10.1103/PhysRevLett.18.546} {\bibfield  {journal} {\bibinfo
  {journal} {Phys. Rev. Lett.}\ }\textbf {\bibinfo {volume} {18}},\ \bibinfo
  {pages} {546} (\bibinfo {year} {1967})}\BibitemShut {NoStop}%
\bibitem [{\citenamefont {Dyakonov}\ and\ \citenamefont
  {Shur}(1993)}]{Dyakonov1993}%
  \BibitemOpen
  \bibfield  {author} {\bibinfo {author} {\bibfnamefont {M.}~\bibnamefont
  {Dyakonov}}\ and\ \bibinfo {author} {\bibfnamefont {M.}~\bibnamefont
  {Shur}},\ }\bibfield  {title} {\bibinfo {title} {Shallow water analogy for a
  ballistic field effect transistor: New mechanism of plasma wave generation by
  dc current},\ }\href {https://doi.org/10.1103/PhysRevLett.71.2465} {\bibfield
   {journal} {\bibinfo  {journal} {Phys. Rev. Lett.}\ }\textbf {\bibinfo
  {volume} {71}},\ \bibinfo {pages} {2465} (\bibinfo {year}
  {1993})}\BibitemShut {NoStop}%
\bibitem [{\citenamefont {Knap}\ \emph {et~al.}(2009)\citenamefont {Knap},
  \citenamefont {Dyakonov}, \citenamefont {Coquillat}, \citenamefont {Teppe},
  \citenamefont {Dyakonova}, \citenamefont {Łusakowski}, \citenamefont
  {Karpierz}, \citenamefont {Sakowicz}, \citenamefont {Valusis}, \citenamefont
  {Seliuta} \emph {et~al.}}]{Knap2009}%
  \BibitemOpen
  \bibfield  {author} {\bibinfo {author} {\bibfnamefont {W.}~\bibnamefont
  {Knap}}, \bibinfo {author} {\bibfnamefont {M.}~\bibnamefont {Dyakonov}},
  \bibinfo {author} {\bibfnamefont {D.}~\bibnamefont {Coquillat}}, \bibinfo
  {author} {\bibfnamefont {F.}~\bibnamefont {Teppe}}, \bibinfo {author}
  {\bibfnamefont {N.}~\bibnamefont {Dyakonova}}, \bibinfo {author}
  {\bibfnamefont {J.}~\bibnamefont {Łusakowski}}, \bibinfo {author}
  {\bibfnamefont {K.}~\bibnamefont {Karpierz}}, \bibinfo {author}
  {\bibfnamefont {M.}~\bibnamefont {Sakowicz}}, \bibinfo {author}
  {\bibfnamefont {G.}~\bibnamefont {Valusis}}, \bibinfo {author} {\bibfnamefont
  {D.}~\bibnamefont {Seliuta}}, \emph {et~al.},\ }\bibfield  {title} {\bibinfo
  {title} {Field effect transistors for terahertz detection: Physics and first
  imaging applications},\ }\href@noop {} {\bibfield  {journal} {\bibinfo
  {journal} {J. Infrared Millim. Terahertz Waves}\ }\textbf {\bibinfo {volume}
  {30}},\ \bibinfo {pages} {1319} (\bibinfo {year} {2009})}\BibitemShut
  {NoStop}%
\bibitem [{\citenamefont {Grigorenko}\ \emph {et~al.}(2012)\citenamefont
  {Grigorenko}, \citenamefont {Polini},\ and\ \citenamefont
  {Novoselov}}]{Grigorenko2012}%
  \BibitemOpen
  \bibfield  {author} {\bibinfo {author} {\bibfnamefont {A.~N.}\ \bibnamefont
  {Grigorenko}}, \bibinfo {author} {\bibfnamefont {M.}~\bibnamefont {Polini}},\
  and\ \bibinfo {author} {\bibfnamefont {K.~S.}\ \bibnamefont {Novoselov}},\
  }\href@noop {} {\bibfield  {journal} {\bibinfo  {journal} {Nat. Photon.}\
  }\textbf {\bibinfo {volume} {6}},\ \bibinfo {pages} {749} (\bibinfo {year}
  {2012})}\BibitemShut {NoStop}%
\bibitem [{\citenamefont {Yao}\ \emph {et~al.}(2018)\citenamefont {Yao},
  \citenamefont {Liu}, \citenamefont {Huang}, \citenamefont {Choi},
  \citenamefont {Xie}, \citenamefont {Flor~Flores}, \citenamefont {Wu},
  \citenamefont {Yu}, \citenamefont {Kwong}, \citenamefont {Huang} \emph
  {et~al.}}]{Yao2018}%
  \BibitemOpen
  \bibfield  {author} {\bibinfo {author} {\bibfnamefont {B.}~\bibnamefont
  {Yao}}, \bibinfo {author} {\bibfnamefont {Y.}~\bibnamefont {Liu}}, \bibinfo
  {author} {\bibfnamefont {S.-W.}\ \bibnamefont {Huang}}, \bibinfo {author}
  {\bibfnamefont {C.}~\bibnamefont {Choi}}, \bibinfo {author} {\bibfnamefont
  {Z.}~\bibnamefont {Xie}}, \bibinfo {author} {\bibfnamefont {J.}~\bibnamefont
  {Flor~Flores}}, \bibinfo {author} {\bibfnamefont {Y.}~\bibnamefont {Wu}},
  \bibinfo {author} {\bibfnamefont {M.}~\bibnamefont {Yu}}, \bibinfo {author}
  {\bibfnamefont {D.-L.}\ \bibnamefont {Kwong}}, \bibinfo {author}
  {\bibfnamefont {Y.}~\bibnamefont {Huang}}, \emph {et~al.},\ }\bibfield
  {title} {\bibinfo {title} {Broadband gate-tunable terahertz plasmons in
  graphene heterostructures},\ }\href {https://doi.org/10.1038/s41566-017-
  0054-7} {\bibfield  {journal} {\bibinfo  {journal} {Nat. Photon.}\ }\textbf
  {\bibinfo {volume} {12}},\ \bibinfo {pages} {22} (\bibinfo {year}
  {2018})}\BibitemShut {NoStop}%
\bibitem [{\citenamefont {Bandurin}\ \emph {et~al.}(2019)\citenamefont
  {Bandurin}, \citenamefont {Svintsov}, \citenamefont {Gayduchenko},
  \citenamefont {Xu}, \citenamefont {Principi}, \citenamefont {Moskotin},
  \citenamefont {Tretyakov}, \citenamefont {Yagodkin}, \citenamefont {Zhukov},
  \citenamefont {Taniguchi} \emph {et~al.}}]{Bandurin2019}%
  \BibitemOpen
  \bibfield  {author} {\bibinfo {author} {\bibfnamefont {D.~A.}\ \bibnamefont
  {Bandurin}}, \bibinfo {author} {\bibfnamefont {D.}~\bibnamefont {Svintsov}},
  \bibinfo {author} {\bibfnamefont {I.}~\bibnamefont {Gayduchenko}}, \bibinfo
  {author} {\bibfnamefont {S.~G.}\ \bibnamefont {Xu}}, \bibinfo {author}
  {\bibfnamefont {A.}~\bibnamefont {Principi}}, \bibinfo {author}
  {\bibfnamefont {M.}~\bibnamefont {Moskotin}}, \bibinfo {author}
  {\bibfnamefont {I.}~\bibnamefont {Tretyakov}}, \bibinfo {author}
  {\bibfnamefont {D.}~\bibnamefont {Yagodkin}}, \bibinfo {author}
  {\bibfnamefont {S.}~\bibnamefont {Zhukov}}, \bibinfo {author} {\bibfnamefont
  {T.}~\bibnamefont {Taniguchi}}, \emph {et~al.},\ }\href@noop {} {\bibfield
  {journal} {\bibinfo  {journal} {Nat. Commun.}\ }\textbf {\bibinfo {volume}
  {9}},\ \bibinfo {pages} {5392} (\bibinfo {year} {2019})}\BibitemShut
  {NoStop}%
\bibitem [{\citenamefont {Mitin}\ \emph {et~al.}(2020)\citenamefont {Mitin},
  \citenamefont {Ryzhii},\ and\ \citenamefont {Otsuji}}]{Mitin2020}%
  \BibitemOpen
  \bibfield  {author} {\bibinfo {author} {\bibfnamefont {V.}~\bibnamefont
  {Mitin}}, \bibinfo {author} {\bibfnamefont {V.}~\bibnamefont {Ryzhii}},\ and\
  \bibinfo {author} {\bibfnamefont {T.}~\bibnamefont {Otsuji}},\ }\href@noop {}
  {\emph {\bibinfo {title} {Graphene-Based Terahertz Electronics and
  Plasmonics: Detector and Emitter Concepts}}}\ (\bibinfo  {publisher} {CRC
  Press},\ \bibinfo {year} {2020})\BibitemShut {NoStop}%
\bibitem [{\citenamefont {Chaplik}(1972)}]{Chaplik1972}%
  \BibitemOpen
  \bibfield  {author} {\bibinfo {author} {\bibfnamefont {A.}~\bibnamefont
  {Chaplik}},\ }\bibfield  {title} {\bibinfo {title} {Possible crystallization
  of charge carriers in low-density inversion layers},\ }\href@noop {}
  {\bibfield  {journal} {\bibinfo  {journal} {Sov. Phys. JETP}\ }\textbf
  {\bibinfo {volume} {35}},\ \bibinfo {pages} {395} (\bibinfo {year}
  {1972})}\BibitemShut {NoStop}%
\bibitem [{\citenamefont {Meziani}\ \emph {et~al.}(2008)\citenamefont
  {Meziani}, \citenamefont {Handa}, \citenamefont {Knap}, \citenamefont
  {Otsuji}, \citenamefont {Sano}, \citenamefont {Popov}, \citenamefont
  {Tsymbalov}, \citenamefont {Coquillat},\ and\ \citenamefont
  {Teppe}}]{Meziani2008}%
  \BibitemOpen
  \bibfield  {author} {\bibinfo {author} {\bibfnamefont {Y.}~\bibnamefont
  {Meziani}}, \bibinfo {author} {\bibfnamefont {H.}~\bibnamefont {Handa}},
  \bibinfo {author} {\bibfnamefont {W.}~\bibnamefont {Knap}}, \bibinfo {author}
  {\bibfnamefont {T.}~\bibnamefont {Otsuji}}, \bibinfo {author} {\bibfnamefont
  {E.}~\bibnamefont {Sano}}, \bibinfo {author} {\bibfnamefont {V.}~\bibnamefont
  {Popov}}, \bibinfo {author} {\bibfnamefont {G.}~\bibnamefont {Tsymbalov}},
  \bibinfo {author} {\bibfnamefont {D.}~\bibnamefont {Coquillat}},\ and\
  \bibinfo {author} {\bibfnamefont {F.}~\bibnamefont {Teppe}},\ }\bibfield
  {title} {\bibinfo {title} {Room temperature terahertz emission from grating
  coupled two-dimensional plasmons},\ }\href
  {https://doi.org/10.1063/1.2919097} {\bibfield  {journal} {\bibinfo
  {journal} {Appl. Phys. Lett.}\ }\textbf {\bibinfo {volume} {92}},\ \bibinfo
  {pages} {201108} (\bibinfo {year} {2008})}\BibitemShut {NoStop}%
\bibitem [{\citenamefont {Nikitin}\ \emph {et~al.}(2016)\citenamefont
  {Nikitin}, \citenamefont {Alonso-González}, \citenamefont {Vélez},
  \citenamefont {Mastel}, \citenamefont {Centeno}, \citenamefont {Pesquera},
  \citenamefont {Zurutuza}, \citenamefont {Casanova}, \citenamefont {Hueso},
  \citenamefont {Koppens} \emph {et~al.}}]{Nikitin2016}%
  \BibitemOpen
  \bibfield  {author} {\bibinfo {author} {\bibfnamefont {A.~Y.}\ \bibnamefont
  {Nikitin}}, \bibinfo {author} {\bibfnamefont {P.}~\bibnamefont
  {Alonso-González}}, \bibinfo {author} {\bibfnamefont {S.}~\bibnamefont
  {Vélez}}, \bibinfo {author} {\bibfnamefont {S.}~\bibnamefont {Mastel}},
  \bibinfo {author} {\bibfnamefont {A.}~\bibnamefont {Centeno}}, \bibinfo
  {author} {\bibfnamefont {A.}~\bibnamefont {Pesquera}}, \bibinfo {author}
  {\bibfnamefont {A.}~\bibnamefont {Zurutuza}}, \bibinfo {author}
  {\bibfnamefont {F.}~\bibnamefont {Casanova}}, \bibinfo {author}
  {\bibfnamefont {L.~E.}\ \bibnamefont {Hueso}}, \bibinfo {author}
  {\bibfnamefont {F.~H.~L.}\ \bibnamefont {Koppens}}, \emph {et~al.},\
  }\bibfield  {title} {\bibinfo {title} {Real-space mapping of tailored sheet
  and edge plasmons in graphene nanoresonators},\ }\href@noop {} {\bibfield
  {journal} {\bibinfo  {journal} {Nat. Photon.}\ }\textbf {\bibinfo {volume}
  {10}},\ \bibinfo {pages} {239} (\bibinfo {year} {2016})}\BibitemShut
  {NoStop}%
\bibitem [{\citenamefont {Zarezin}\ \emph {et~al.}(2021)\citenamefont
  {Zarezin}, \citenamefont {Gusikhin}, \citenamefont {Andreev}, \citenamefont
  {Muravev},\ and\ \citenamefont {Kukushkin}}]{Zarezin2021}%
  \BibitemOpen
  \bibfield  {author} {\bibinfo {author} {\bibfnamefont {A.}~\bibnamefont
  {Zarezin}}, \bibinfo {author} {\bibfnamefont {P.}~\bibnamefont {Gusikhin}},
  \bibinfo {author} {\bibfnamefont {I.}~\bibnamefont {Andreev}}, \bibinfo
  {author} {\bibfnamefont {V.}~\bibnamefont {Muravev}},\ and\ \bibinfo {author}
  {\bibfnamefont {I.}~\bibnamefont {Kukushkin}},\ }\bibfield  {title} {\bibinfo
  {title} {Plasmon excitations in partially screened two-dimensional electron
  systems (brief review)},\ }\href {https://doi.org/10.1134/S0021364021110096}
  {\bibfield  {journal} {\bibinfo  {journal} {JETP Lett.}\ }\textbf {\bibinfo
  {volume} {113}},\ \bibinfo {pages} {713} (\bibinfo {year}
  {2021})}\BibitemShut {NoStop}%
\bibitem [{\citenamefont {Alcaraz~Iranzo}\ \emph {et~al.}(2018)\citenamefont
  {Alcaraz~Iranzo}, \citenamefont {Nanot}, \citenamefont {Dias}, \citenamefont
  {Epstein}, \citenamefont {Peng}, \citenamefont {Efetov}, \citenamefont
  {Lundeberg}, \citenamefont {Parret}, \citenamefont {Osmond}, \citenamefont
  {Hong} \emph {et~al.}}]{Alcaraz2018}%
  \BibitemOpen
  \bibfield  {author} {\bibinfo {author} {\bibfnamefont {D.}~\bibnamefont
  {Alcaraz~Iranzo}}, \bibinfo {author} {\bibfnamefont {S.}~\bibnamefont
  {Nanot}}, \bibinfo {author} {\bibfnamefont {E.~J.}\ \bibnamefont {Dias}},
  \bibinfo {author} {\bibfnamefont {I.}~\bibnamefont {Epstein}}, \bibinfo
  {author} {\bibfnamefont {C.}~\bibnamefont {Peng}}, \bibinfo {author}
  {\bibfnamefont {D.~K.}\ \bibnamefont {Efetov}}, \bibinfo {author}
  {\bibfnamefont {M.~B.}\ \bibnamefont {Lundeberg}}, \bibinfo {author}
  {\bibfnamefont {R.}~\bibnamefont {Parret}}, \bibinfo {author} {\bibfnamefont
  {J.}~\bibnamefont {Osmond}}, \bibinfo {author} {\bibfnamefont {J.-Y.}\
  \bibnamefont {Hong}}, \emph {et~al.},\ }\bibfield  {title} {\bibinfo {title}
  {Probing the ultimate plasmon confinement limits with a van der waals
  heterostructure},\ }\href {https://doi.org/10.1126/science.aar8438}
  {\bibfield  {journal} {\bibinfo  {journal} {Science}\ }\textbf {\bibinfo
  {volume} {360}},\ \bibinfo {pages} {291} (\bibinfo {year}
  {2018})}\BibitemShut {NoStop}%
\bibitem [{\citenamefont {Epstein}\ \emph {et~al.}(2020)\citenamefont
  {Epstein}, \citenamefont {Alcaraz}, \citenamefont {Huang}, \citenamefont
  {Pusapati}, \citenamefont {Hugonin}, \citenamefont {Kumar}, \citenamefont
  {Deputy}, \citenamefont {Khodkov}, \citenamefont {Rappoport}, \citenamefont
  {Hong} \emph {et~al.}}]{Epstein2020}%
  \BibitemOpen
  \bibfield  {author} {\bibinfo {author} {\bibfnamefont {I.}~\bibnamefont
  {Epstein}}, \bibinfo {author} {\bibfnamefont {D.}~\bibnamefont {Alcaraz}},
  \bibinfo {author} {\bibfnamefont {Z.}~\bibnamefont {Huang}}, \bibinfo
  {author} {\bibfnamefont {V.-V.}\ \bibnamefont {Pusapati}}, \bibinfo {author}
  {\bibfnamefont {J.-P.}\ \bibnamefont {Hugonin}}, \bibinfo {author}
  {\bibfnamefont {A.}~\bibnamefont {Kumar}}, \bibinfo {author} {\bibfnamefont
  {X.~M.}\ \bibnamefont {Deputy}}, \bibinfo {author} {\bibfnamefont
  {T.}~\bibnamefont {Khodkov}}, \bibinfo {author} {\bibfnamefont {T.~G.}\
  \bibnamefont {Rappoport}}, \bibinfo {author} {\bibfnamefont {J.-Y.}\
  \bibnamefont {Hong}}, \emph {et~al.},\ }\bibfield  {title} {\bibinfo {title}
  {Far-field excitation of single graphene plasmon cavities with
  ultracompressed mode volumes},\ }\href
  {https://doi.org/10.1126/science.abb1570} {\bibfield  {journal} {\bibinfo
  {journal} {Science}\ }\textbf {\bibinfo {volume} {368}},\ \bibinfo {pages}
  {1219} (\bibinfo {year} {2020})}\BibitemShut {NoStop}%
\bibitem [{\citenamefont {Mylnikov}\ and\ \citenamefont
  {Svintsov}(2022)}]{Mylnikov2022}%
  \BibitemOpen
  \bibfield  {author} {\bibinfo {author} {\bibfnamefont {D.}~\bibnamefont
  {Mylnikov}}\ and\ \bibinfo {author} {\bibfnamefont {D.}~\bibnamefont
  {Svintsov}},\ }\bibfield  {title} {\bibinfo {title} {Limiting capabilities of
  two-dimensional plasmonics in electromagnetic wave detection},\ }\href
  {https://doi.org/10.1103/PhysRevApplied.17.064055} {\bibfield  {journal}
  {\bibinfo  {journal} {Phys. Rev. Appl.}\ }\textbf {\bibinfo {volume} {17}},\
  \bibinfo {pages} {064055} (\bibinfo {year} {2022})}\BibitemShut {NoStop}%
\bibitem [{\citenamefont {Glattli}\ \emph {et~al.}(1985)\citenamefont
  {Glattli}, \citenamefont {Andrei}, \citenamefont {Deville}, \citenamefont
  {Poitrenaud},\ and\ \citenamefont {Williams}}]{Glattli1985}%
  \BibitemOpen
  \bibfield  {author} {\bibinfo {author} {\bibfnamefont {D.~C.}\ \bibnamefont
  {Glattli}}, \bibinfo {author} {\bibfnamefont {E.~Y.}\ \bibnamefont {Andrei}},
  \bibinfo {author} {\bibfnamefont {G.}~\bibnamefont {Deville}}, \bibinfo
  {author} {\bibfnamefont {J.}~\bibnamefont {Poitrenaud}},\ and\ \bibinfo
  {author} {\bibfnamefont {F.~I.~B.}\ \bibnamefont {Williams}},\ }\href@noop {}
  {\bibfield  {journal} {\bibinfo  {journal} {Phys. Rev. Lett.}\ }\textbf
  {\bibinfo {volume} {54}},\ \bibinfo {pages} {1710} (\bibinfo {year}
  {1985})}\BibitemShut {NoStop}%
\bibitem [{\citenamefont {Fetter}(1986)}]{Fetter1986}%
  \BibitemOpen
  \bibfield  {author} {\bibinfo {author} {\bibfnamefont {A.~L.}\ \bibnamefont
  {Fetter}},\ }\bibfield  {title} {\bibinfo {title} {Magnetoplasmons in a
  two-dimensional electron fluid: Disk geometry},\ }\href@noop {} {\bibfield
  {journal} {\bibinfo  {journal} {Phys. Rev. B}\ }\textbf {\bibinfo {volume}
  {33}},\ \bibinfo {pages} {5221} (\bibinfo {year} {1986})}\BibitemShut
  {NoStop}%
\bibitem [{\citenamefont {Zabolotnykh}\ and\ \citenamefont
  {Volkov}(2021)}]{Zabolotnykh2021}%
  \BibitemOpen
  \bibfield  {author} {\bibinfo {author} {\bibfnamefont {A.~A.}\ \bibnamefont
  {Zabolotnykh}}\ and\ \bibinfo {author} {\bibfnamefont {V.~A.}\ \bibnamefont
  {Volkov}},\ }\bibfield  {title} {\bibinfo {title} {Electrically controllable
  cyclotron resonance},\ }\href {https://doi.org/10.1103/PhysRevB.103.125301}
  {\bibfield  {journal} {\bibinfo  {journal} {Phys. Rev. B}\ }\textbf {\bibinfo
  {volume} {103}},\ \bibinfo {pages} {125301} (\bibinfo {year}
  {2021})}\BibitemShut {NoStop}%
\bibitem [{\citenamefont {Gusikhin}\ \emph {et~al.}(2018)\citenamefont
  {Gusikhin}, \citenamefont {Muravev}, \citenamefont {Zagitova},\ and\
  \citenamefont {Kukushkin}}]{Gusikhin2018}%
  \BibitemOpen
  \bibfield  {author} {\bibinfo {author} {\bibfnamefont {P.~A.}\ \bibnamefont
  {Gusikhin}}, \bibinfo {author} {\bibfnamefont {V.~M.}\ \bibnamefont
  {Muravev}}, \bibinfo {author} {\bibfnamefont {A.~A.}\ \bibnamefont
  {Zagitova}},\ and\ \bibinfo {author} {\bibfnamefont {I.~V.}\ \bibnamefont
  {Kukushkin}},\ }\bibfield  {title} {\bibinfo {title} {Drastic reduction of
  plasmon damping in two-dimensional electron disks},\ }\href
  {https://doi.org/10.1103/PhysRevLett.121.176804} {\bibfield  {journal}
  {\bibinfo  {journal} {Phys. Rev. Lett.}\ }\textbf {\bibinfo {volume} {121}},\
  \bibinfo {pages} {176804} (\bibinfo {year} {2018})}\BibitemShut {NoStop}%
\bibitem [{\citenamefont {Mikhailov}(2004)}]{Mikhailov2004}%
  \BibitemOpen
  \bibfield  {author} {\bibinfo {author} {\bibfnamefont {S.~A.}\ \bibnamefont
  {Mikhailov}},\ }\bibfield  {title} {\bibinfo {title} {Microwave-induced
  magnetotransport phenomena in two-dimensional electron systems: Importance of
  electrodynamic effects},\ }\href@noop {} {\bibfield  {journal} {\bibinfo
  {journal} {Phys. Rev. B}\ }\textbf {\bibinfo {volume} {70}},\ \bibinfo
  {pages} {165311} (\bibinfo {year} {2004})}\BibitemShut {NoStop}%
\bibitem [{\citenamefont {Chung}\ \emph {et~al.}(2021)\citenamefont {Chung},
  \citenamefont {Villegas~Rosales}, \citenamefont {Baldwin}, \citenamefont
  {Madathil}, \citenamefont {West}, \citenamefont {Shayegan},\ and\
  \citenamefont {Pfeiffer}}]{Chung2021}%
  \BibitemOpen
  \bibfield  {author} {\bibinfo {author} {\bibfnamefont {Y.~J.}\ \bibnamefont
  {Chung}}, \bibinfo {author} {\bibfnamefont {K.}~\bibnamefont
  {Villegas~Rosales}}, \bibinfo {author} {\bibfnamefont {K.}~\bibnamefont
  {Baldwin}}, \bibinfo {author} {\bibfnamefont {P.}~\bibnamefont {Madathil}},
  \bibinfo {author} {\bibfnamefont {K.}~\bibnamefont {West}}, \bibinfo {author}
  {\bibfnamefont {M.}~\bibnamefont {Shayegan}},\ and\ \bibinfo {author}
  {\bibfnamefont {L.}~\bibnamefont {Pfeiffer}},\ }\bibfield  {title} {\bibinfo
  {title} {Ultra-high-quality two-dimensional electron systems},\ }\href
  {https://doi.org/10.1038/s41563-021-00942-3} {\bibfield  {journal} {\bibinfo
  {journal} {Nat. Mater.}\ }\textbf {\bibinfo {volume} {20}},\ \bibinfo {pages}
  {632} (\bibinfo {year} {2021})}\BibitemShut {NoStop}%
\bibitem [{\citenamefont {Batygin}\ and\ \citenamefont
  {Toptygin}(1964)}]{Batygin1964}%
  \BibitemOpen
  \bibfield  {author} {\bibinfo {author} {\bibfnamefont {V.~V.}\ \bibnamefont
  {Batygin}}\ and\ \bibinfo {author} {\bibfnamefont {I.~N.}\ \bibnamefont
  {Toptygin}},\ }\href@noop {} {\emph {\bibinfo {title} {Problems in
  Electrodynamics}}}\ (\bibinfo  {publisher} {Academic Press},\ \bibinfo {year}
  {1964})\ Chap.~\bibinfo {chapter} {13}\BibitemShut {NoStop}%
\bibitem [{\citenamefont {Li}\ and\ \citenamefont {Arnoldus}(2010)}]{Li2010}%
  \BibitemOpen
  \bibfield  {author} {\bibinfo {author} {\bibfnamefont {X.}~\bibnamefont
  {Li}}\ and\ \bibinfo {author} {\bibfnamefont {H.~F.}\ \bibnamefont
  {Arnoldus}},\ }\bibfield  {title} {\bibinfo {title} {Electric dipole
  radiation near a mirror},\ }\href
  {https://doi.org/10.1103/PhysRevA.81.053844} {\bibfield  {journal} {\bibinfo
  {journal} {Phys. Rev. A}\ }\textbf {\bibinfo {volume} {81}},\ \bibinfo
  {pages} {053844} (\bibinfo {year} {2010})}\BibitemShut {NoStop}%
\bibitem [{\citenamefont {Haroche}\ and\ \citenamefont
  {Kleppner}(1989)}]{Haroche1989}%
  \BibitemOpen
  \bibfield  {author} {\bibinfo {author} {\bibfnamefont {S.}~\bibnamefont
  {Haroche}}\ and\ \bibinfo {author} {\bibfnamefont {D.}~\bibnamefont
  {Kleppner}},\ }\bibfield  {title} {\bibinfo {title} {Cavity quantum
  electrodynamics},\ }\href {https://doi.org/10.1063/1.881201} {\bibfield
  {journal} {\bibinfo  {journal} {Phys. Today}\ }\textbf {\bibinfo {volume}
  {42}},\ \bibinfo {pages} {24} (\bibinfo {year} {1989})}\BibitemShut {NoStop}%
\bibitem [{\citenamefont {Villegas~Rosales}\ \emph {et~al.}(2022)\citenamefont
  {Villegas~Rosales}, \citenamefont {Madathil}, \citenamefont {Chung},
  \citenamefont {Pfeiffer}, \citenamefont {West}, \citenamefont {Baldwin},\
  and\ \citenamefont {Shayegan}}]{Rosales2022}%
  \BibitemOpen
  \bibfield  {author} {\bibinfo {author} {\bibfnamefont {K.~A.}\ \bibnamefont
  {Villegas~Rosales}}, \bibinfo {author} {\bibfnamefont {P.~T.}\ \bibnamefont
  {Madathil}}, \bibinfo {author} {\bibfnamefont {Y.~J.}\ \bibnamefont {Chung}},
  \bibinfo {author} {\bibfnamefont {L.~N.}\ \bibnamefont {Pfeiffer}}, \bibinfo
  {author} {\bibfnamefont {K.~W.}\ \bibnamefont {West}}, \bibinfo {author}
  {\bibfnamefont {K.~W.}\ \bibnamefont {Baldwin}},\ and\ \bibinfo {author}
  {\bibfnamefont {M.}~\bibnamefont {Shayegan}},\ }\bibfield  {title} {\bibinfo
  {title} {Composite fermion mass: Experimental measurements in ultrahigh
  quality two-dimensional electron systems},\ }\href
  {https://doi.org/10.1103/PhysRevB.106.L041301} {\bibfield  {journal}
  {\bibinfo  {journal} {Phys. Rev. B}\ }\textbf {\bibinfo {volume} {106}},\
  \bibinfo {pages} {L041301} (\bibinfo {year} {2022})}\BibitemShut {NoStop}%
\bibitem [{\citenamefont {Muravev}\ \emph {et~al.}(2017)\citenamefont
  {Muravev}, \citenamefont {Andreev}, \citenamefont {Belyanin}, \citenamefont
  {Gubarev},\ and\ \citenamefont {Kukushkin}}]{Muravev2017}%
  \BibitemOpen
  \bibfield  {author} {\bibinfo {author} {\bibfnamefont {V.~M.}\ \bibnamefont
  {Muravev}}, \bibinfo {author} {\bibfnamefont {I.~V.}\ \bibnamefont
  {Andreev}}, \bibinfo {author} {\bibfnamefont {V.~N.}\ \bibnamefont
  {Belyanin}}, \bibinfo {author} {\bibfnamefont {S.~I.}\ \bibnamefont
  {Gubarev}},\ and\ \bibinfo {author} {\bibfnamefont {I.~V.}\ \bibnamefont
  {Kukushkin}},\ }\bibfield  {title} {\bibinfo {title} {Observation of
  axisymmetric dark plasma excitations in a two-dimensional electron system},\
  }\href@noop {} {\bibfield  {journal} {\bibinfo  {journal} {Phys. Rev. B}\
  }\textbf {\bibinfo {volume} {96}},\ \bibinfo {pages} {045421} (\bibinfo
  {year} {2017})}\BibitemShut {NoStop}%
\bibitem [{\citenamefont {Zoric}\ \emph {et~al.}(2011)\citenamefont {Zoric},
  \citenamefont {Zach}, \citenamefont {Kasemo},\ and\ \citenamefont
  {Langhammer}}]{Zoric2011}%
  \BibitemOpen
  \bibfield  {author} {\bibinfo {author} {\bibfnamefont {I.}~\bibnamefont
  {Zoric}}, \bibinfo {author} {\bibfnamefont {M.}~\bibnamefont {Zach}},
  \bibinfo {author} {\bibfnamefont {B.}~\bibnamefont {Kasemo}},\ and\ \bibinfo
  {author} {\bibfnamefont {C.}~\bibnamefont {Langhammer}},\ }\bibfield  {title}
  {\bibinfo {title} {Gold, platinum, and aluminum nanodisk plasmons: material
  independence, subradiance, and damping mechanisms},\ }\href@noop {}
  {\bibfield  {journal} {\bibinfo  {journal} {ACS Nano}\ }\textbf {\bibinfo
  {volume} {5}},\ \bibinfo {pages} {2535} (\bibinfo {year} {2011})}\BibitemShut
  {NoStop}%
\bibitem [{\citenamefont {Andreev}\ \emph {et~al.}(2014)\citenamefont
  {Andreev}, \citenamefont {Muravev}, \citenamefont {Belyanin},\ and\
  \citenamefont {Kukushkin}}]{Andreev2014}%
  \BibitemOpen
  \bibfield  {author} {\bibinfo {author} {\bibfnamefont {I.~V.}\ \bibnamefont
  {Andreev}}, \bibinfo {author} {\bibfnamefont {V.~M.}\ \bibnamefont
  {Muravev}}, \bibinfo {author} {\bibfnamefont {V.~N.}\ \bibnamefont
  {Belyanin}},\ and\ \bibinfo {author} {\bibfnamefont {I.~V.}\ \bibnamefont
  {Kukushkin}},\ }\bibfield  {title} {\bibinfo {title} {Measurement of
  cyclotron resonance relaxation time in the two-dimensional electron system},\
  }\href@noop {} {\bibfield  {journal} {\bibinfo  {journal} {Appl. Phys.
  Lett.}\ }\textbf {\bibinfo {volume} {105}},\ \bibinfo {pages} {202106}
  (\bibinfo {year} {2014})}\BibitemShut {NoStop}%
\bibitem [{\citenamefont {Andreev}\ \emph {et~al.}(2015)\citenamefont
  {Andreev}, \citenamefont {Muravev}, \citenamefont {Belyanin},\ and\
  \citenamefont {Kukushkin}}]{Andreev2015}%
  \BibitemOpen
  \bibfield  {author} {\bibinfo {author} {\bibfnamefont {I.~V.}\ \bibnamefont
  {Andreev}}, \bibinfo {author} {\bibfnamefont {V.~M.}\ \bibnamefont
  {Muravev}}, \bibinfo {author} {\bibfnamefont {V.~N.}\ \bibnamefont
  {Belyanin}},\ and\ \bibinfo {author} {\bibfnamefont {I.~V.}\ \bibnamefont
  {Kukushkin}},\ }\href@noop {} {\bibfield  {journal} {\bibinfo  {journal}
  {JETP Lett.}\ }\textbf {\bibinfo {volume} {102}},\ \bibinfo {pages} {821}
  (\bibinfo {year} {2015})}\BibitemShut {NoStop}%
\bibitem [{\citenamefont {Andreev}\ \emph {et~al.}(2021)\citenamefont
  {Andreev}, \citenamefont {Muravev}, \citenamefont {Semenov},\ and\
  \citenamefont {Kukushkin}}]{Andreev2021}%
  \BibitemOpen
  \bibfield  {author} {\bibinfo {author} {\bibfnamefont {I.~V.}\ \bibnamefont
  {Andreev}}, \bibinfo {author} {\bibfnamefont {V.~M.}\ \bibnamefont
  {Muravev}}, \bibinfo {author} {\bibfnamefont {N.~D.}\ \bibnamefont
  {Semenov}},\ and\ \bibinfo {author} {\bibfnamefont {I.~V.}\ \bibnamefont
  {Kukushkin}},\ }\bibfield  {title} {\bibinfo {title} {Observation of acoustic
  plasma waves with a velocity approaching the speed of light},\ }\href
  {https://doi.org/10.1103/PhysRevB.103.115420} {\bibfield  {journal} {\bibinfo
   {journal} {Phys. Rev. B}\ }\textbf {\bibinfo {volume} {103}},\ \bibinfo
  {pages} {115420} (\bibinfo {year} {2021})}\BibitemShut {NoStop}%
\bibitem [{\citenamefont {Kukushkin}\ \emph {et~al.}(1989)\citenamefont
  {Kukushkin}, \citenamefont {von Klitzing}, \citenamefont {Ploog},
  \citenamefont {Kirpichev},\ and\ \citenamefont {Shepel}}]{Kukushkin1989}%
  \BibitemOpen
  \bibfield  {author} {\bibinfo {author} {\bibfnamefont {I.~V.}\ \bibnamefont
  {Kukushkin}}, \bibinfo {author} {\bibfnamefont {K.}~\bibnamefont {von
  Klitzing}}, \bibinfo {author} {\bibfnamefont {K.}~\bibnamefont {Ploog}},
  \bibinfo {author} {\bibfnamefont {V.~E.}\ \bibnamefont {Kirpichev}},\ and\
  \bibinfo {author} {\bibfnamefont {B.~N.}\ \bibnamefont {Shepel}},\ }\bibfield
   {title} {\bibinfo {title} {Reduction of the electron density in
  $\mathrm{GaAs}-\mathrm{Al}_{x}\mathrm{Ga}_{1-x}\mathrm{As}$ single
  heterojunctions by continuous photoexcitation},\ }\href@noop {} {\bibfield
  {journal} {\bibinfo  {journal} {Phys. Rev. B}\ }\textbf {\bibinfo {volume}
  {40}},\ \bibinfo {pages} {4179} (\bibinfo {year} {1989})}\BibitemShut
  {NoStop}%
\bibitem [{\citenamefont {Zagorodnev}\ \emph {et~al.}(2021)\citenamefont
  {Zagorodnev}, \citenamefont {Rodionov},\ and\ \citenamefont
  {Zabolotnykh}}]{Zagorodnev2021}%
  \BibitemOpen
  \bibfield  {author} {\bibinfo {author} {\bibfnamefont {I.~V.}\ \bibnamefont
  {Zagorodnev}}, \bibinfo {author} {\bibfnamefont {D.~A.}\ \bibnamefont
  {Rodionov}},\ and\ \bibinfo {author} {\bibfnamefont {A.~A.}\ \bibnamefont
  {Zabolotnykh}},\ }\bibfield  {title} {\bibinfo {title} {Effect of retardation
  on the frequency and linewidth of plasma resonances in a two-dimensional disk
  of electron gas},\ }\href@noop {} {\bibfield  {journal} {\bibinfo  {journal}
  {Phys. Rev. B}\ }\textbf {\bibinfo {volume} {103}},\ \bibinfo {pages}
  {195431} (\bibinfo {year} {2021})}\BibitemShut {NoStop}%
\bibitem [{\citenamefont {Abramowitz}\ \emph {et~al.}(1988)\citenamefont
  {Abramowitz}, \citenamefont {Stegun},\ and\ \citenamefont
  {Romer}}]{Abramowitz1988}%
  \BibitemOpen
  \bibfield  {author} {\bibinfo {author} {\bibfnamefont {M.}~\bibnamefont
  {Abramowitz}}, \bibinfo {author} {\bibfnamefont {I.~A.}\ \bibnamefont
  {Stegun}},\ and\ \bibinfo {author} {\bibfnamefont {R.~H.}\ \bibnamefont
  {Romer}},\ }\href@noop {} {\emph {\bibinfo {title} {Handbook of mathematical
  functions with formulas, graphs, and mathematical tables}}}\ (\bibinfo
  {publisher} {American Association of Physics Teachers},\ \bibinfo {year}
  {1988})\ Chap.~\bibinfo {chapter} {9}\BibitemShut {NoStop}%
\bibitem [{\citenamefont {Duan}\ \emph {et~al.}(2022)\citenamefont {Duan},
  \citenamefont {Alfaro-Mozaz}, \citenamefont {Taboada-Guti{\'e}rrez},
  \citenamefont {Dolado}, \citenamefont {{\'A}lvarez-P{\'e}rez}, \citenamefont
  {Titova}, \citenamefont {Bylinkin}, \citenamefont {Tresguerres-Mata},
  \citenamefont {Mart{\'\i}n-S{\'a}nchez}, \citenamefont {Liu} \emph
  {et~al.}}]{Duan2022}%
  \BibitemOpen
  \bibfield  {author} {\bibinfo {author} {\bibfnamefont {J.}~\bibnamefont
  {Duan}}, \bibinfo {author} {\bibfnamefont {F.~J.}\ \bibnamefont
  {Alfaro-Mozaz}}, \bibinfo {author} {\bibfnamefont {J.}~\bibnamefont
  {Taboada-Guti{\'e}rrez}}, \bibinfo {author} {\bibfnamefont {I.}~\bibnamefont
  {Dolado}}, \bibinfo {author} {\bibfnamefont {G.}~\bibnamefont
  {{\'A}lvarez-P{\'e}rez}}, \bibinfo {author} {\bibfnamefont {E.}~\bibnamefont
  {Titova}}, \bibinfo {author} {\bibfnamefont {A.}~\bibnamefont {Bylinkin}},
  \bibinfo {author} {\bibfnamefont {A.~I.~F.}\ \bibnamefont
  {Tresguerres-Mata}}, \bibinfo {author} {\bibfnamefont {J.}~\bibnamefont
  {Mart{\'\i}n-S{\'a}nchez}}, \bibinfo {author} {\bibfnamefont
  {S.}~\bibnamefont {Liu}}, \emph {et~al.},\ }\bibfield  {title} {\bibinfo
  {title} {Active and passive tuning of ultranarrow resonances in polaritonic
  nanoantennas},\ }\href {https://doi.org/10.1002/adma.202104954} {\bibfield
  {journal} {\bibinfo  {journal} {Adv. Mater.}\ }\textbf {\bibinfo {volume}
  {34}},\ \bibinfo {pages} {2104954} (\bibinfo {year} {2022})}\BibitemShut
  {NoStop}%
\bibitem [{\citenamefont {Landau}\ and\ \citenamefont
  {Lifshitz}(1975)}]{Landau1975}%
  \BibitemOpen
  \bibfield  {author} {\bibinfo {author} {\bibfnamefont {L.~D.}\ \bibnamefont
  {Landau}}\ and\ \bibinfo {author} {\bibfnamefont {E.~M.}\ \bibnamefont
  {Lifshitz}},\ }\href@noop {} {\emph {\bibinfo {title} {The Classical Theory
  of Fields}}}\ (\bibinfo  {publisher} {Butterworth-Heinemann},\ \bibinfo
  {year} {1975})\ Chap.~\bibinfo {chapter} {9}\BibitemShut {NoStop}%
\bibitem [{Note1()}]{Note1}%
  \BibitemOpen
  \bibinfo {note} {Wolfram Mathematica code to carry out the analysis is
  publicly available at
  https://github.com/danilrodionov/gated-disk.git}\BibitemShut {NoStop}%
\bibitem [{\citenamefont {Tretyakov}(2014)}]{Tretyakov2014}%
  \BibitemOpen
  \bibfield  {author} {\bibinfo {author} {\bibfnamefont {S.}~\bibnamefont
  {Tretyakov}},\ }\bibfield  {title} {\bibinfo {title} {Maximizing absorption
  and scattering by dipole particles},\ }\href
  {https://doi.org/10.1007/s11468-014-9699-y} {\bibfield  {journal} {\bibinfo
  {journal} {Plasmonics}\ }\textbf {\bibinfo {volume} {9}},\ \bibinfo {pages}
  {935} (\bibinfo {year} {2014})}\BibitemShut {NoStop}%
\bibitem [{\citenamefont {Zagorodnev}\ \emph {et~al.}(2019)\citenamefont
  {Zagorodnev}, \citenamefont {Rodionov}, \citenamefont {Zabolotnykh},\ and\
  \citenamefont {Volkov}}]{Zagorodnev2019}%
  \BibitemOpen
  \bibfield  {author} {\bibinfo {author} {\bibfnamefont {I.~V.}\ \bibnamefont
  {Zagorodnev}}, \bibinfo {author} {\bibfnamefont {D.~A.}\ \bibnamefont
  {Rodionov}}, \bibinfo {author} {\bibfnamefont {A.~A.}\ \bibnamefont
  {Zabolotnykh}},\ and\ \bibinfo {author} {\bibfnamefont {V.~A.}\ \bibnamefont
  {Volkov}},\ }\bibfield  {title} {\bibinfo {title} {Microwave absorption by
  axisymmetric plasmon mode in 2d electron disk},\ }\href@noop {} {\bibfield
  {journal} {\bibinfo  {journal} {Semiconductors}\ }\textbf {\bibinfo {volume}
  {53}},\ \bibinfo {pages} {1873} (\bibinfo {year} {2019})}\BibitemShut
  {NoStop}%
\bibitem [{\citenamefont {Falson}\ \emph {et~al.}(2016)\citenamefont {Falson},
  \citenamefont {Kozuka}, \citenamefont {Uchida}, \citenamefont {Smet},
  \citenamefont {Arima}, \citenamefont {Tsukazaki},\ and\ \citenamefont
  {Kawasaki}}]{Falson2016}%
  \BibitemOpen
  \bibfield  {author} {\bibinfo {author} {\bibfnamefont {J.}~\bibnamefont
  {Falson}}, \bibinfo {author} {\bibfnamefont {Y.}~\bibnamefont {Kozuka}},
  \bibinfo {author} {\bibfnamefont {M.}~\bibnamefont {Uchida}}, \bibinfo
  {author} {\bibfnamefont {J.~H.}\ \bibnamefont {Smet}}, \bibinfo {author}
  {\bibfnamefont {T.-h.}\ \bibnamefont {Arima}}, \bibinfo {author}
  {\bibfnamefont {A.}~\bibnamefont {Tsukazaki}},\ and\ \bibinfo {author}
  {\bibfnamefont {M.}~\bibnamefont {Kawasaki}},\ }\bibfield  {title} {\bibinfo
  {title} {Mgzno/zno heterostructures with electron mobility exceeding $1\times
  10^{6}$ cm$^2$/$\mathrm{Vs}$},\ }\href {https://doi.org/10.1038/srep26598}
  {\bibfield  {journal} {\bibinfo  {journal} {Sci. Rep.}\ }\textbf {\bibinfo
  {volume} {6}},\ \bibinfo {pages} {1} (\bibinfo {year} {2016})}\BibitemShut
  {NoStop}%
\bibitem [{\citenamefont {Tschirky}\ \emph {et~al.}(2017)\citenamefont
  {Tschirky}, \citenamefont {Mueller}, \citenamefont {Lehner}, \citenamefont
  {F\"alt}, \citenamefont {Ihn}, \citenamefont {Ensslin},\ and\ \citenamefont
  {Wegscheider}}]{Tschirky2017}%
  \BibitemOpen
  \bibfield  {author} {\bibinfo {author} {\bibfnamefont {T.}~\bibnamefont
  {Tschirky}}, \bibinfo {author} {\bibfnamefont {S.}~\bibnamefont {Mueller}},
  \bibinfo {author} {\bibfnamefont {C.~A.}\ \bibnamefont {Lehner}}, \bibinfo
  {author} {\bibfnamefont {S.}~\bibnamefont {F\"alt}}, \bibinfo {author}
  {\bibfnamefont {T.}~\bibnamefont {Ihn}}, \bibinfo {author} {\bibfnamefont
  {K.}~\bibnamefont {Ensslin}},\ and\ \bibinfo {author} {\bibfnamefont
  {W.}~\bibnamefont {Wegscheider}},\ }\bibfield  {title} {\bibinfo {title}
  {Scattering mechanisms of highest-mobility
  $\mathrm{InAs}/\mathrm{Al}_{x}\mathrm{Ga}_{1\ensuremath{-}x}\mathrm{Sb}$
  quantum wells},\ }\href {https://doi.org/10.1103/PhysRevB.95.115304}
  {\bibfield  {journal} {\bibinfo  {journal} {Phys. Rev. B}\ }\textbf {\bibinfo
  {volume} {95}},\ \bibinfo {pages} {115304} (\bibinfo {year}
  {2017})}\BibitemShut {NoStop}%
\end{thebibliography}%

\end{document}